\documentclass[aps,preprint,superscriptaddress]{revtex4}
\usepackage{amsfonts}
\usepackage{mathrsfs}
\usepackage{graphicx}
\usepackage{amsmath}
\usepackage{amssymb}
\usepackage{bbm}
\usepackage{subfigure}
\usepackage{caption}
\usepackage{float}
\usepackage{epsfig}
\usepackage{graphicx}
\usepackage{tikz}
\usepackage{color}
\usepackage{physics}
\usepackage{indentfirst}
\usepackage{hyperref}
\usepackage{setspace}
\usepackage{braket}
\usepackage{slashed}
\usepackage{multirow}
\usepackage{diagbox}
\usepackage{cancel}
\usepackage{caption}
\usepackage{array}
\newcommand{\dslah}{\slash \! \! \! }

\allowdisplaybreaks[3]

\begin{document}

\title{The Spectra of $p\bar\Lambda$ and $p\bar\Sigma$ Hexaquark States}

\author{Xuan-Heng Zhang}
\email{zhangxuanheng22@mails.ucas.ac.cn}
\author{Sheng-Qi Zhang}
\email{zhangshengqi20@mails.ucas.ac.cn}
\author{Cong-Feng Qiao}
\email{qiaocf@ucas.ac.cn, corresponding author}

\affiliation{School of Physical Sciences, University of Chinese Academy of Sciences, Yuquan Road 19A, Beijing 100049, China\\ \vspace{2.7cm}}

\begin{abstract}

Motivated by the recent observation of the $J^P = 1^+$ resonance $X(2085)$ in the $p\bar{\Lambda}$ system by the BESIII collaboration, we study the molecular states of $p\bar{\Lambda}$ and $p\bar{\Sigma}$ with baryon-antibaryon structures within the framework of the QCD sum rules. Non-perturbative contributions up to dimension 13 in quark operator expansion (OPE) are considered. Our calculation results indicate the possible existence of 6 $p\bar{\Lambda}$ and $p\bar{\Sigma}$ molecular states with quantum numbers $J^{P}=0^{-}, 0^{+}, 1^{-}$, which do not support $X(2085)$ being a $p\bar{\Lambda}$ or $p\bar{\Sigma}$ molecular state, or at least not its main component. Additionally, the previously observed state $X(2075)$ also lies in the vicinity of $p\bar{\Lambda}$ and $p\bar{\Sigma}$ molecular states, and its inner structure might not be entirely transparent, rather possibly a mixture of tetraquark and hexaquark states. The possible decay modes of the concerned states are analyzed.
\end{abstract}
\vspace{1cm}

\maketitle

\newpage
\topmargin -1.5cm

\section{Introduction}
In the 1960s, to explain new particles discovered in large colliders, Gell-Mann et al. proposed the quark model (QM) \cite{Gell-Mann:1964ewy,Zweig:1964jf}, marking the beginning of human exploration of strong interactions. By 1973, with the establishment of quantum chromodynamics (QCD), the theory describing strong interactions was essentially complete. QCD calculations are fundamentally performed at the level of quarks and gluons. However, no free quarks or gluons have been observed experimentally so far; only hadrons, which are composed of quarks and gluons, can be detected. Therefore, studying the properties of hadrons can help us understand strong interactions and enrich our knowledge of the microscopic world. 

Though conventional hadronic states include mesons $(q\bar q)$ and baryons $(qqq)$, other configurations, such as hybrid states, glueballs and multiquark states more than three quarks, are also allowed in the QM. In 2003, the Belle Collaboration discovered the state $X(3872)$ \cite{Belle:2003nnu}, which is widely believed to be an exotic one, i.e. a mutiquark state \cite{Pakvasa:2003ea,Voloshin:2003nt,Liu:2007uj,Matheus:2006xi,Lee:2008uy,Chen:2013pya}. Since then, with the development of experimental technology, numerous exotic state candidates have been observed in experiments \cite{LHCb:2015yax,BESIII:2013ris,Belle:2014nuw,LHCb:2020bwg,BESIIICollaboration:2022kwh,BESIII:2022riz,BESIII:2022iwi}, especially tetraquark states and pentaquark states \cite{LHCb:2015yax,Chen:2015moa,Wang:2015epa}. Research on the properties of these hadronic states has garnered a widespread attention in past two decades.

There are various theoretical methods for studying exotic states, including quark models \cite{Gell-Mann:1964ewy,Zweig:1964jf}, the MIT bag model \cite{Chodos:1974je}, lattice QCD \cite{Wilson:1974sk}, chiral effective theory \cite{tHooft:1976snw}, AdS/QCD \cite{Erlich:2005qh}, and QCD sum rules (QCDSR) \cite{Shifman:1978bx,Shifman:1978by}, and others. Among these, the QCDSR method proposed by Shifman, Vainshtein, and Zakharov (SVZ) is based on the first principles of QCD and effectively incorporates the non-perturbative contributions of QCD. QCDSR is a well-established and effective approach for investigating hadron properties, providing analytical expressions for hadron masses \cite{Shifman:1978bx,Shifman:1978by,Zhang:2022obn,Wan:2020fsk,Wan:2020oxt}, form factors \cite{Ioffe:1982qb,Wang:2007ys,Wu:2024gcq,Zhao:2020mod,Neishabouri:2024gbc,Aliev:2010uy,Zhang:2023nxl,Zhang:2024ick}, decay constants \cite{Dominguez:1987ea,Khodjamirian:2000ds,Gelhausen:2013wia,Hu:2017dzi,Narison:2020wql}, and other observables \cite{Colangelo:2000dp}. Numerous studies have demonstrated that QCDSR provides reliable results in investigating various properties of exotic states\cite{Colangelo:2000dp,Wang:2025sic}. As an example, for the first discovered exotic state $X(3872)$, QCDSR has successfully provided explanations for its structure \cite{Matheus:2006xi,Lee:2008uy,Chen:2013pya}.

In fact, some kind of hexaquark states exists in nature, such as the dibaryons, a bound state of two baryons. However, the existence of baryon-antibaryon states, or hexaquark states with equal amount of quarks and antiquarks, has not been demonstrated. Although many theoretical predictions have been made \cite{Jaffe:1976yi,Mulders:1980vx,Balachandran:1983dj,Sakai:1999qm,Ikeda:2007nz,Bashkanov:2013cla,Shanahan:2011su,Clement:2016vnl}, these have yet to be observed experimentally \cite{BaBar:2018hpv}. The molecular hexaquark states are comparatively simple in structure, especially the baryonium. The $\Lambda_c\bar{\Lambda}_c$ heavy baryonium has been used to explain the structure of $Y(4260)$ \cite{BaBar:2005hhc,Qiao:2005av,Qiao:2007ce}. Subsequently, a series of theoretical analyses on light baryonium states have been performed in QCDSR \cite{Wan:2021vny,Wang:2006sna}, which can also successfully explain the structures of $X(1835)$ \cite{BES:2003aic} and $X(1840)$ \cite{BESIII:2023vvr} observed in experiments. Furthermore, QCDSR has also provided predictions for the mass spectra of other baryonium states \cite{Wan:2019ake,Chen:2016ymy,Wang:2021qmn}.

Molecular hexaquark states would not be limited to baryonium; they could be composed of a baryon and a different antibaryon. Experimentally, several structures have been observed in the final state of $p\bar{\Lambda}$, including $X(2075)$ \cite{BES:2004fgd} and the recently discovered $X(2085)$ \cite{BESIII:2023kgz}, both of which lie close to the threshold of $p\bar{\Lambda}$. Among them, the quantum number of $X(2085)$ has been determined to be $J^P=1^+$. Theoretically, $X(2075)$ and $X(2085)$ could be interpreted as a baryon-antibaryon molecular state of $p\bar{\Lambda}$, or have large component of $p\bar{\Lambda}$. Nevertheless, it is tempting to consider other structures near the threshold and with the same quantum number, for instance the molecular state $p\bar{\Sigma}$ \cite{Huang:2011zq}, which may enrich our understanding of newly observed states of $X(2075)$ and $X(2085)$. The molecular states $p\bar{\Lambda}$ and $p\bar{\Sigma}$ belong to the flavor SU(3) nonet \cite{Yuan:2005pf}, and studies based on the quark models have already provided preliminary calculations of their mass spectra \cite{Ding:2005tr,Huang:2011zq}.

In this work, we will determine the mass spectra of the ground molecular states $p\bar{\Lambda}$ and $p\bar{\Sigma}$ in the framework of QCDSR. The results will be compared to the experimental measurements, and potential decay modes will be analyzed. The structure of this paper is organized as follows: in Sect.\ref{form}, we briefly introduce the theoretical framework of QCDSR and present the fundamental formulas used in our calculations. In Sect.\ref{na}, we provide numerical analyses and results. Section \ref{dm} discusses the possible decay modes of $p\bar{\Lambda}$ and $p\bar{\Sigma}$. Finally, in Sect.\ref{dc}, we compare our results with experimental observations and summarize our findings.

\section{\label{form}Formalism}
\subsection{Choices of the Currents}
To calculate the mass spectrum of $p\bar{\Lambda}$ and $p\bar{\Sigma}$ in the framework of QCDSR, it is essential to first select appropriate hadronic interpolating currents. There are two independent interpolating currents for the baryon octet\cite{Chen:2008qv}, and the other interpolating current structures can be obtained by linear combinations of them through Fierz transformations \cite{Fierz:1937wjm}. In our calculation, the masses of quarks $u$ and $d$ are rather smaller than that of $s$-quark, so we take the limit $m_u = m_d \to 0$, which simplifies the structure of the interpolating currents. The two simplified interpolating currents we have chosen are  \cite{Chung:1981cc,Chung:1981wm}
\begin{equation}\label{eta1}
\text{Type-1:}\quad \eta_{1\mathscr{B}}(x) = \varepsilon_{abc}\left[q_a^{i T}(x) \mathcal{C} q_b^j(x)\right] \gamma_5 q_c^k(x)\ ,
\end{equation}
\begin{equation}\label{eta2}
\text{Type-2:}\quad \eta_{2\mathscr{B}}(x) = \varepsilon_{abc}\left[q_a^{i T}(x) \mathcal{C} \gamma_5 q_b^j(x)\right] q_c^k(x)\ ,
\end{equation}
where for $p, \Lambda, \Sigma$, the indices $(i,j,k)$ take the following values: $(u,d,u)$ for $p$, $(u,d,s)$ for $\Lambda$, and $(u,s,d)$ for $\Sigma$. $a,b,c$ are the color indices. $\mathscr{B}$ denotes an arbitrary light baryon.

For the baryon-antibaryon type hexaquark molecular states $\mathscr{B}\bar{\mathscr{B^{\prime}}}$, the quantum numbers of the ground states will take $J^{P} = 0^{-}, 0^{+}, 1^{-}, 1^{+}$, and the corresponding interpolating current structures are given by
\begin{align}
    j^{0^{-}}(x) &= \mathrm{i} \bar{\eta}_{\mathscr{B^{\prime}}}(x) \gamma_5 \eta_{\mathscr{B}}(x),\label{j0-} \\
    j^{0^{+}}(x) &= \bar{\eta}_{\mathscr{B^{\prime}}}(x) \eta_{\mathscr{B}}(x),\label{j0+} \\
    j_\mu^{1^{-}}(x) &= \bar{\eta}_{\mathscr{B^{\prime}}}(x) \gamma_\mu \eta_{\mathscr{B}}(x),\label{j1-} \\
    j_\mu^{1^{+}}(x) &=  \bar{\eta}_{\mathscr{B^{\prime}}}(x) \gamma_\mu \gamma_5 \eta_{\mathscr{B}}(x)\label{j1+},
\end{align}
where the two baryonic interpolating currents $\bar{\eta}_{\mathscr{B^{\prime}}}(x)$ and $\eta_{\mathscr{B}}(x)$ for each molecular state must be chosen uniformly from either Eqs.\eqref{eta1} or \eqref{eta2}. Thus, for each baryonic interpolating current, there exist four possible molecular state structures of hexaquark, corresponding to four different quantum numbers $J^{P}$.

\subsection{Two-point Correlation Functions}
After selecting the interpolating currents Eqs.\eqref{j0-}--\eqref{j1+}, we can evaluate the two-point correlation functions of the interpolating currents
\begin{equation}\label{Pi1}
    \Pi(q^2) = \mathrm{i} \int \dd^4 x \, \mathrm{e}^{\mathrm{i} q \cdot x} \bra{\Omega} \mathbb{T}\{ j(x), j^{\dagger}(0) \} \ket{\Omega},
\end{equation}
\begin{equation}\label{Pi2}
    \Pi_{\mu\nu}(q^2) = \mathrm{i} \int \dd^4 x \, \mathrm{e}^{\mathrm{i} q \cdot x} \bra{\Omega} \mathbb{T}\{ j_{\mu}(x), j_{\nu}^{\dagger}(0) \} \ket{\Omega},
\end{equation}
where $j(x), j_{\mu}(x)$ are the interpolating currents corresponding to the molecular states with $J = 0, 1$, respectively, and $\ket{\Omega}$ represents the physical QCD vacuum state. The tensor-type correlation function  $\Pi_{\mu\nu}$ cannot fully represent the $J=1$ molecular state. It can be written as the sum of two parts
\begin{equation}
    \Pi_{\mu\nu}(q^2) = -\left( g_{\mu\nu} - \frac{q_{\mu} q_{\nu}}{q^2} \right) \Pi_1(q^2) + \frac{q_{\mu} q_{\nu}}{q^2} \Pi_0(q^2),
\end{equation}
where the subscripts $1$ and $0$ correspond to particles with spin $1$ and $0$, respectively. By projecting, we obtain
\begin{equation}
    \Pi_1(q^2) = -\frac{1}{3} \left( g^{\mu\nu} - \frac{q^{\mu} q^{\nu}}{q^2} \right) \Pi_{\mu\nu}(q^2),
\end{equation}
which is the two-point correlation function for the $J = 1$ molecular state current.

\subsubsection{OPE Side}
We can analytically evaluate the correlation functions given in Eqs.\eqref{Pi1}-\eqref{Pi2}. For the Type-1 baryon current Eq.\eqref{eta1}, the correlation functions in coordinate space are denoted as  
\begin{equation}\label{1pLambda0}
\begin{aligned}
     \Pi^{\text{Type-1}}_{J=0}(q^2)=&\int[\mathcal{D}x]\Tr\left[\mathcal{S}^{cc_1}_{q_1}(-x)\left(\mathbbm{1}, \mathrm{i}\gamma_5\right)\mathcal{S}^{f_1f}_{u}(x)\left(\mathbbm{1}, \mathrm{i}\gamma_5\right)\right]\times\\
			&\Tr\left[\mathcal{C}\mathcal{S}^{Taa_1}_{u}(-x)\mathcal{C}\mathcal{S}^{bb_1}_{q_2}(-x)\right]\times\Tr\left[\mathcal{C}\mathcal{S}^{Td_1d}_{u}(x)\mathcal{C}\mathcal{S}^{e_1e}_{d}(x)\right],
\end{aligned}
\end{equation}
\begin{equation}
\begin{aligned}
     \Pi^{\text{Type-1}}_{J=1,\mu\nu}(q^2)=&\int[\mathcal{D}x]\Tr\left[\mathcal{S}^{cc_1}_{q_1}(-x)\gamma_{\mu}\left(\mathbbm{1},\gamma_5\right)\mathcal{S}^{f_1f}_{u}(x)\gamma_{\nu}\left(\mathbbm{1}, \gamma_5\right)\right]\times\\
			&\Tr\left[\mathcal{C}\mathcal{S}^{Taa_1}_{u}(-x)\mathcal{C}\mathcal{S}^{bb_1}_{q_2}(-x)\right]\times\Tr\left[\mathcal{C}\mathcal{S}^{Td_1d}_{u}(x)\mathcal{C}\mathcal{S}^{e_1e}_{d}(x)\right],
\end{aligned}
\end{equation}
where $a, b, c, a_1, b_1$ and $c_1$ are color indices, and $\mathbbm{1}$ and $\mathrm{i}\gamma_5$ correspond to the quantum numbers $0^{+}(1^{-})$ and $0^{-}(1^+)$, respectively.  For $p\bar\Lambda$, $q_1=s,\ q_2=d$ and for $p\bar\Sigma$, $q_1=d,\ q_2=s$.  We have redefined the integral measure $\int[\mathcal{D}x]=-\mathrm{i}\int\dd^4 x\ \mathrm{e}^{\mathrm{i}q\cdot x}\varepsilon_{abc}\varepsilon_{a_1b_1c_1}\varepsilon_{def}\varepsilon_{d_1e_1f_1}$. 

For the Type-2 baryon current Eq.\eqref{eta2}, the correlation functions are similarly given by
\begin{equation}
\begin{aligned}
     \Pi^{\text{Type-2}}_{J=0}(x)=&\int[\mathcal{D}x]\Tr\left[\mathcal{S}^{cc_1}_{q_1}(-x)\left(\mathbbm{1},\mathrm{i}\gamma_5\right)\mathcal{S}^{f_1f}_{u}(x)\left(\mathbbm{1},\mathrm{i}\gamma_5\right)\right]\times\\
			&\Tr\left[\mathcal{C}\mathcal{S}^{Taa_1}_{u}(-x)\mathcal{C}\gamma_5\mathcal{S}^{bb_1}_{q_2}(-x)\gamma_5\right]\times\Tr\left[\mathcal{C}\mathcal{S}^{Td_1d}_{u}(x)\mathcal{C}\gamma_5\mathcal{S}^{e_1e}_{d}(x)\gamma_5\right],
\end{aligned}
\end{equation}
\begin{equation}\label{2psigma1}
\begin{aligned}
     \Pi^{\text{Type-2}}_{J=1,\mu\nu}(x)=&\int[\mathcal{D}x]\Tr\left[\mathcal{S}^{cc_1}_{q_1}(-x)\gamma_{\mu}\left(\mathbbm{1},\mathrm{i}\gamma_5\right)\mathcal{S}^{f_1f}_{u}(x)\gamma_{\nu}\left(\mathbbm{1},\mathrm{i}\gamma_5\right)\right]\times\\
			&\Tr\left[\mathcal{C}\mathcal{S}^{Taa_1}_{u}(-x)\mathcal{C}\gamma_5\mathcal{S}^{bb_1}_{q_2}(-x)\gamma_5\right]\times\Tr\left[\mathcal{C}\mathcal{S}^{Td_1d}_{u}(x)\mathcal{C}\gamma_5\mathcal{S}^{e_1e}_{d}(x)\gamma_5\right].
\end{aligned}
\end{equation}

The propagator $\mathcal{S}^{ab}_{q}(x)$ in the above correlation functions represents the full QCD propagator, which incorporates both perturbative and non-perturbative contributions at all orders of vacuum condensates. The full propagator for a light quark $q = u, d, s$ can be expressed as 
\begin{equation}
    \begin{aligned}
        \mathrm{i}\mathcal{S}_{q}^{jk}(x) =& \mathrm{i} \delta^{jk} \frac{\dslah{x}}{2\pi^2 x^4} 
- \delta^{jk} m_q \frac{1}{4\pi^2 x^2} 
- \mathrm{i} t^{jk}_a \frac{G^a_{\alpha\beta}}{32 \pi^2 x^2} 
\left( \sigma^{\alpha\beta} \dslah{x} + \dslah{x} \sigma^{\alpha\beta} \right) 
- \delta^{jk} \frac{\langle \bar{q} q \rangle}{12} 
+ \mathrm{i} \delta^{jk} \frac{\dslah{x}}{48} m_q \langle \bar{q} q \rangle 
\\
&- \delta^{jk} \frac{x^2}{192} \langle g_s \bar{q} \sigma \cdot G q \rangle + \mathrm{i} \delta^{jk} \frac{x^2 \dslah{x}}{1152} m_q \langle g_s \bar{q} \sigma \cdot G q \rangle 
- t^{jk}_a \frac{\sigma_{\alpha\beta}}{192} \langle g_s \bar{q} \sigma \cdot G q \rangle \\ 
&- \mathrm{i} t^{jk}_a \frac{1}{768} 
\left(\sigma_{\alpha\beta} \dslah{x} + \dslah{x} \sigma_{\alpha\beta} \right) 
m_q \langle g_s \bar{q} \sigma \cdot G q \rangle.
    \end{aligned}
\end{equation}
More details on the full propagator can be found in Refs.\cite{Wang:2013vex,Albuquerque:2012jbz}. The typical Feynman diagrams of the correlation functions are shown in Fig.\ref{Feymann}, which also includes the contributions of the non-perturbative vacuum condensates.
\begin{figure}[htpb]
    \centering
\tikzset{every picture/.style={line width=0.75pt}} 
\begin{tikzpicture}[x=0.75pt,y=0.75pt,yscale=-0.8,xscale=0.8]
\draw   (17.92,124.77) .. controls (17.92,91.79) and (44.71,65.05) .. (77.75,65.05) .. controls (110.79,65.05) and (137.57,91.79) .. (137.57,124.77) .. controls (137.57,157.75) and (110.79,184.49) .. (77.75,184.49) .. controls (44.71,184.49) and (17.92,157.75) .. (17.92,124.77) -- cycle ;
\draw   (17.92,124.77) .. controls (17.92,101.43) and (44.71,82.51) .. (77.75,82.51) .. controls (110.79,82.51) and (137.57,101.43) .. (137.57,124.77) .. controls (137.57,148.11) and (110.79,167.03) .. (77.75,167.03) .. controls (44.71,167.03) and (17.92,148.11) .. (17.92,124.77) -- cycle ;
\draw   (17.92,124.77) .. controls (17.92,112.21) and (44.71,102.03) .. (77.75,102.03) .. controls (110.79,102.03) and (137.57,112.21) .. (137.57,124.77) .. controls (137.57,137.33) and (110.79,147.51) .. (77.75,147.51) .. controls (44.71,147.51) and (17.92,137.33) .. (17.92,124.77) -- cycle ;
\draw  [line width=1.5]  (76.83,61.37) -- (84.19,65.51) -- (76.83,69.64) ;
\draw  [line width=1.5]  (76.83,78.83) -- (84.19,82.97) -- (76.83,87.1) ;
\draw  [line width=1.5]  (75.91,98.13) -- (83.27,102.26) -- (75.91,106.39) ;
\draw  [line width=1.5]  (83.59,151.11) -- (75.92,147.57) -- (82.93,142.87) ;
\draw  [line width=1.5]  (84.51,170.4) -- (76.84,166.87) -- (83.85,162.16) ;
\draw  [line width=1.5]  (85.43,188.78) -- (77.76,185.24) -- (84.77,180.54) ;
\draw   (182.79,124.77) .. controls (182.79,91.79) and (209.58,65.05) .. (242.62,65.05) .. controls (275.66,65.05) and (302.45,91.79) .. (302.45,124.77) .. controls (302.45,157.75) and (275.66,184.49) .. (242.62,184.49) .. controls (209.58,184.49) and (182.79,157.75) .. (182.79,124.77) -- cycle ;
\draw   (182.79,124.77) .. controls (182.79,101.43) and (209.58,82.51) .. (242.62,82.51) .. controls (275.66,82.51) and (302.45,101.43) .. (302.45,124.77) .. controls (302.45,148.11) and (275.66,167.03) .. (242.62,167.03) .. controls (209.58,167.03) and (182.79,148.11) .. (182.79,124.77) -- cycle ;
\draw  [line width=1.5]  (241.7,61.37) -- (249.06,65.51) -- (241.7,69.64) ;
\draw  [line width=1.5]  (241.7,78.83) -- (249.06,82.97) -- (241.7,87.1) ;
\draw  [line width=1.5]  (248.46,151.11) -- (240.79,147.57) -- (247.8,142.87) ;
\draw  [line width=1.5]  (249.38,170.4) -- (241.71,166.87) -- (248.72,162.16) ;
\draw  [line width=1.5]  (250.3,188.78) -- (242.63,185.24) -- (249.64,180.54) ;
\draw  [draw opacity=0] (301.37,127.53) .. controls (298.15,138.89) and (273.26,148.02) .. (242.92,148.42) .. controls (210.56,148.85) and (184.14,139.2) .. (183.53,126.79) -- (242.62,125.69) -- cycle ; \draw   (301.37,127.53) .. controls (298.15,138.89) and (273.26,148.02) .. (242.92,148.42) .. controls (210.56,148.85) and (184.14,139.2) .. (183.53,126.79) ;  
\draw  [fill={rgb, 255:red, 0; green, 0; blue, 0 }  ,fill opacity=1 ] (257.32,102.74) .. controls (257.32,101.73) and (258.14,100.91) .. (259.16,100.91) .. controls (260.17,100.91) and (261,101.73) .. (261,102.74) .. controls (261,103.76) and (260.17,104.58) .. (259.16,104.58) .. controls (258.14,104.58) and (257.32,103.76) .. (257.32,102.74) -- cycle ;
\draw    (183.54,126.53) -- (230.63,102.74) ;
\draw    (301.37,127.52) -- (258.24,102.74) ;
\draw   (83.91,283.48) .. controls (85.22,283.82) and (86.54,285.2) .. (86.55,287.95) .. controls (86.57,293.47) and (81.33,293.48) .. (81.32,290.73) .. controls (81.31,287.97) and (86.55,287.95) .. (86.57,293.47) .. controls (86.59,298.98) and (81.35,299) .. (81.34,296.24) .. controls (81.33,293.48) and (86.57,293.47) .. (86.59,298.98) .. controls (86.59,300.19) and (86.34,301.14) .. (85.95,301.85) ;
\draw   (514,124.66) .. controls (514,91.68) and (540.78,64.94) .. (573.83,64.94) .. controls (606.87,64.94) and (633.65,91.68) .. (633.65,124.66) .. controls (633.65,157.64) and (606.87,184.38) .. (573.83,184.38) .. controls (540.78,184.38) and (514,157.64) .. (514,124.66) -- cycle ;
\draw   (514,124.66) .. controls (514,101.32) and (540.78,82.39) .. (573.83,82.39) .. controls (606.87,82.39) and (633.65,101.32) .. (633.65,124.66) .. controls (633.65,148) and (606.87,166.92) .. (573.83,166.92) .. controls (540.78,166.92) and (514,148) .. (514,124.66) -- cycle ;
\draw   (514,124.66) .. controls (514,112.1) and (540.78,101.92) .. (573.83,101.92) .. controls (606.87,101.92) and (633.65,112.1) .. (633.65,124.66) .. controls (633.65,137.22) and (606.87,147.4) .. (573.83,147.4) .. controls (540.78,147.4) and (514,137.22) .. (514,124.66) -- cycle ;
\draw  [line width=1.5]  (572.91,61.26) -- (580.27,65.4) -- (572.91,69.53) ;
\draw  [line width=1.5]  (572.91,78.72) -- (580.27,82.85) -- (572.91,86.99) ;
\draw  [line width=1.5]  (571.98,98.01) -- (579.35,102.15) -- (571.98,106.28) ;
\draw  [line width=1.5]  (579.66,151) -- (572,147.46) -- (579.01,142.76) ;
\draw  [line width=1.5]  (580.59,170.29) -- (572.92,166.75) -- (579.93,162.05) ;
\draw  [line width=1.5]  (581.51,188.67) -- (573.84,185.13) -- (580.85,180.43) ;
\draw  [fill={rgb, 255:red, 0; green, 0; blue, 0 }  ,fill opacity=1 ] (226.94,103.66) .. controls (226.94,102.65) and (227.77,101.83) .. (228.78,101.83) .. controls (229.8,101.83) and (230.63,102.65) .. (230.63,103.66) .. controls (230.63,104.68) and (229.8,105.5) .. (228.78,105.5) .. controls (227.77,105.5) and (226.94,104.68) .. (226.94,103.66) -- cycle ;
\draw   (23.4,325.74) .. controls (23.4,292.76) and (50.18,266.02) .. (83.22,266.02) .. controls (116.26,266.02) and (143.05,292.76) .. (143.05,325.74) .. controls (143.05,358.72) and (116.26,385.46) .. (83.22,385.46) .. controls (50.18,385.46) and (23.4,358.72) .. (23.4,325.74) -- cycle ;
\draw   (23.4,325.74) .. controls (23.4,302.4) and (50.18,283.48) .. (83.22,283.48) .. controls (116.26,283.48) and (143.05,302.4) .. (143.05,325.74) .. controls (143.05,349.08) and (116.26,368) .. (83.22,368) .. controls (50.18,368) and (23.4,349.08) .. (23.4,325.74) -- cycle ;
\draw  [line width=1.5]  (82.3,262.34) -- (89.67,266.48) -- (82.3,270.61) ;
\draw  [line width=1.5]  (82.3,279.8) -- (89.67,283.94) -- (82.3,288.07) ;
\draw  [line width=1.5]  (89.06,352.08) -- (81.39,348.54) -- (88.4,343.84) ;
\draw  [line width=1.5]  (89.98,371.37) -- (82.31,367.84) -- (89.32,363.13) ;
\draw  [line width=1.5]  (90.9,389.75) -- (83.23,386.21) -- (90.25,381.51) ;
\draw  [draw opacity=0] (141.97,328.49) .. controls (138.75,339.86) and (113.87,348.99) .. (83.52,349.39) .. controls (51.16,349.82) and (24.74,340.17) .. (24.13,327.76) -- (83.22,326.66) -- cycle ; \draw   (141.97,328.49) .. controls (138.75,339.86) and (113.87,348.99) .. (83.52,349.39) .. controls (51.16,349.82) and (24.74,340.17) .. (24.13,327.76) ;  
\draw  [fill={rgb, 255:red, 0; green, 0; blue, 0 }  ,fill opacity=1 ] (97.92,303.71) .. controls (97.92,302.7) and (98.74,301.88) .. (99.76,301.88) .. controls (100.78,301.88) and (101.6,302.7) .. (101.6,303.71) .. controls (101.6,304.73) and (100.78,305.55) .. (99.76,305.55) .. controls (98.74,305.55) and (97.92,304.73) .. (97.92,303.71) -- cycle ;
\draw    (24.15,327.5) -- (71.23,303.71) ;
\draw    (141.97,328.49) -- (98.84,303.71) ;
\draw  [fill={rgb, 255:red, 0; green, 0; blue, 0 }  ,fill opacity=1 ] (67.55,304.63) .. controls (67.55,303.62) and (68.37,302.79) .. (69.39,302.79) .. controls (70.4,302.79) and (71.23,303.62) .. (71.23,304.63) .. controls (71.23,305.65) and (70.4,306.47) .. (69.39,306.47) .. controls (68.37,306.47) and (67.55,305.65) .. (67.55,304.63) -- cycle ;
\draw  [fill={rgb, 255:red, 0; green, 0; blue, 0 }  ,fill opacity=1 ] (83.19,303.71) .. controls (83.19,302.7) and (84.02,301.88) .. (85.03,301.88) .. controls (86.05,301.88) and (86.87,302.7) .. (86.87,303.71) .. controls (86.87,304.73) and (86.05,305.55) .. (85.03,305.55) .. controls (84.02,305.55) and (83.19,304.73) .. (83.19,303.71) -- cycle ;
\draw   (575.02,103.84) .. controls (576.57,104.07) and (578.13,105.21) .. (578.13,107.51) .. controls (578.15,112.1) and (571.95,112.12) .. (571.94,109.37) .. controls (571.93,106.61) and (578.13,106.59) .. (578.15,111.18) .. controls (578.16,115.78) and (571.96,115.8) .. (571.95,113.04) .. controls (571.94,110.29) and (578.14,110.26) .. (578.16,114.86) .. controls (578.17,116.65) and (577.23,117.74) .. (576.09,118.25) ;
\draw  [fill={rgb, 255:red, 0; green, 0; blue, 0 }  ,fill opacity=1 ] (574.72,118.25) .. controls (574.72,117.24) and (575.54,116.41) .. (576.56,116.41) .. controls (577.57,116.41) and (578.4,117.24) .. (578.4,118.25) .. controls (578.4,119.27) and (577.57,120.09) .. (576.56,120.09) .. controls (575.54,120.09) and (574.72,119.27) .. (574.72,118.25) -- cycle ;
\draw   (575.93,145.8) .. controls (574.39,145.52) and (572.87,144.33) .. (572.94,142.03) .. controls (573.09,137.44) and (579.29,137.64) .. (579.2,140.39) .. controls (579.12,143.15) and (572.91,142.95) .. (573.06,138.36) .. controls (573.2,133.77) and (579.41,133.96) .. (579.32,136.72) .. controls (579.23,139.47) and (573.03,139.28) .. (573.18,134.69) .. controls (573.23,132.9) and (574.21,131.84) .. (575.37,131.37) ;
\draw  [fill={rgb, 255:red, 0; green, 0; blue, 0 }  ,fill opacity=1 ] (577.66,132.33) .. controls (577.62,133.35) and (576.77,134.14) .. (575.75,134.11) .. controls (574.74,134.07) and (573.94,133.22) .. (573.98,132.2) .. controls (574.02,131.19) and (574.87,130.4) .. (575.88,130.43) .. controls (576.9,130.47) and (577.69,131.32) .. (577.66,132.33) -- cycle ;
\draw   (187.92,326.77) .. controls (187.92,293.79) and (214.71,267.05) .. (247.75,267.05) .. controls (280.79,267.05) and (307.57,293.79) .. (307.57,326.77) .. controls (307.57,359.75) and (280.79,386.49) .. (247.75,386.49) .. controls (214.71,386.49) and (187.92,359.75) .. (187.92,326.77) -- cycle ;
\draw   (187.92,326.77) .. controls (187.92,303.43) and (214.71,284.51) .. (247.75,284.51) .. controls (280.79,284.51) and (307.57,303.43) .. (307.57,326.77) .. controls (307.57,350.11) and (280.79,369.03) .. (247.75,369.03) .. controls (214.71,369.03) and (187.92,350.11) .. (187.92,326.77) -- cycle ;
\draw   (187.92,326.77) .. controls (187.92,314.21) and (214.71,304.03) .. (247.75,304.03) .. controls (280.79,304.03) and (307.57,314.21) .. (307.57,326.77) .. controls (307.57,339.33) and (280.79,349.51) .. (247.75,349.51) .. controls (214.71,349.51) and (187.92,339.33) .. (187.92,326.77) -- cycle ;
\draw  [line width=1.5]  (246.83,263.37) -- (254.19,267.51) -- (246.83,271.64) ;
\draw  [line width=1.5]  (246.83,280.83) -- (254.19,284.97) -- (246.83,289.1) ;
\draw  [line width=1.5]  (245.91,300.13) -- (253.27,304.26) -- (245.91,308.39) ;
\draw  [line width=1.5]  (253.59,353.11) -- (245.92,349.57) -- (252.93,344.87) ;
\draw  [line width=1.5]  (254.51,372.4) -- (246.84,368.87) -- (253.85,364.16) ;
\draw  [line width=1.5]  (255.43,390.78) -- (247.76,387.24) -- (254.77,382.54) ;
\draw   (214.22,375.84) .. controls (215.87,376.07) and (217.53,377.21) .. (217.54,379.51) .. controls (217.55,384.1) and (210.95,384.12) .. (210.94,381.37) .. controls (210.93,378.61) and (217.53,378.59) .. (217.55,383.18) .. controls (217.57,387.78) and (210.96,387.8) .. (210.95,385.04) .. controls (210.94,382.29) and (217.55,382.26) .. (217.56,386.86) .. controls (217.58,391.45) and (210.97,391.47) .. (210.96,388.72) .. controls (210.95,385.96) and (217.56,385.94) .. (217.58,390.53) .. controls (217.59,395.13) and (210.99,395.15) .. (210.98,392.39) .. controls (210.97,389.64) and (217.57,389.61) .. (217.59,394.21) .. controls (217.61,398.8) and (211,398.82) .. (210.99,396.07) .. controls (210.98,393.31) and (217.59,393.29) .. (217.6,397.88) .. controls (217.62,402.48) and (211.01,402.5) .. (211,399.74) .. controls (210.99,396.99) and (217.6,396.96) .. (217.62,401.56) .. controls (217.62,401.89) and (217.58,402.19) .. (217.52,402.48) ;
\draw  [fill={rgb, 255:red, 0; green, 0; blue, 0 }  ,fill opacity=1 ] (214.72,403.25) .. controls (214.72,402.24) and (215.54,401.41) .. (216.56,401.41) .. controls (217.57,401.41) and (218.4,402.24) .. (218.4,403.25) .. controls (218.4,404.27) and (217.57,405.09) .. (216.56,405.09) .. controls (215.54,405.09) and (214.72,404.27) .. (214.72,403.25) -- cycle ;
\draw   (281.27,375.75) .. controls (279.46,375.99) and (277.65,377.14) .. (277.65,379.44) .. controls (277.67,384.03) and (284.93,384.01) .. (284.92,381.25) .. controls (284.91,378.49) and (277.65,378.52) .. (277.67,383.11) .. controls (277.68,387.71) and (284.94,387.68) .. (284.93,384.93) .. controls (284.92,382.17) and (277.66,382.2) .. (277.68,386.79) .. controls (277.7,391.38) and (284.95,391.36) .. (284.94,388.6) .. controls (284.93,385.84) and (277.68,385.87) .. (277.69,390.46) .. controls (277.71,395.06) and (284.97,395.03) .. (284.96,392.28) .. controls (284.95,389.52) and (277.69,389.55) .. (277.71,394.14) .. controls (277.72,398.73) and (284.98,398.71) .. (284.97,395.95) .. controls (284.96,393.19) and (277.7,393.22) .. (277.72,397.81) .. controls (277.74,402.41) and (285,402.38) .. (284.99,399.63) .. controls (284.98,396.87) and (277.72,396.9) .. (277.73,401.49) .. controls (277.74,402.4) and (278.03,403.13) .. (278.48,403.69) ;
\draw  [fill={rgb, 255:red, 0; green, 0; blue, 0 }  ,fill opacity=1 ] (277.72,404.25) .. controls (277.72,403.24) and (278.54,402.41) .. (279.56,402.41) .. controls (280.57,402.41) and (281.4,403.24) .. (281.4,404.25) .. controls (281.4,405.27) and (280.57,406.09) .. (279.56,406.09) .. controls (278.54,406.09) and (277.72,405.27) .. (277.72,404.25) -- cycle ;
\draw   (353.79,327.77) .. controls (353.79,294.79) and (380.58,268.05) .. (413.62,268.05) .. controls (446.66,268.05) and (473.45,294.79) .. (473.45,327.77) .. controls (473.45,360.75) and (446.66,387.49) .. (413.62,387.49) .. controls (380.58,387.49) and (353.79,360.75) .. (353.79,327.77) -- cycle ;
\draw   (353.79,327.77) .. controls (353.79,304.43) and (380.58,285.51) .. (413.62,285.51) .. controls (446.66,285.51) and (473.45,304.43) .. (473.45,327.77) .. controls (473.45,351.11) and (446.66,370.03) .. (413.62,370.03) .. controls (380.58,370.03) and (353.79,351.11) .. (353.79,327.77) -- cycle ;
\draw  [line width=1.5]  (412.7,264.37) -- (420.06,268.51) -- (412.7,272.64) ;
\draw  [line width=1.5]  (412.7,281.83) -- (420.06,285.97) -- (412.7,290.1) ;
\draw  [line width=1.5]  (419.46,354.11) -- (411.79,350.57) -- (418.8,345.87) ;
\draw  [line width=1.5]  (420.38,373.4) -- (412.71,369.87) -- (419.72,365.16) ;
\draw  [line width=1.5]  (421.3,391.78) -- (413.63,388.24) -- (420.64,383.54) ;
\draw  [draw opacity=0] (472.37,330.53) .. controls (469.15,341.89) and (444.26,351.02) .. (413.92,351.42) .. controls (381.56,351.85) and (355.14,342.2) .. (354.53,329.79) -- (413.62,328.69) -- cycle ; \draw   (472.37,330.53) .. controls (469.15,341.89) and (444.26,351.02) .. (413.92,351.42) .. controls (381.56,351.85) and (355.14,342.2) .. (354.53,329.79) ;  
\draw  [fill={rgb, 255:red, 0; green, 0; blue, 0 }  ,fill opacity=1 ] (428.32,305.74) .. controls (428.32,304.73) and (429.14,303.91) .. (430.16,303.91) .. controls (431.17,303.91) and (432,304.73) .. (432,305.74) .. controls (432,306.76) and (431.17,307.58) .. (430.16,307.58) .. controls (429.14,307.58) and (428.32,306.76) .. (428.32,305.74) -- cycle ;
\draw    (354.54,329.53) -- (401.63,305.74) ;
\draw    (472.37,330.52) -- (429.24,305.74) ;
\draw  [fill={rgb, 255:red, 0; green, 0; blue, 0 }  ,fill opacity=1 ] (397.94,306.66) .. controls (397.94,305.65) and (398.77,304.83) .. (399.78,304.83) .. controls (400.8,304.83) and (401.63,305.65) .. (401.63,306.66) .. controls (401.63,307.68) and (400.8,308.5) .. (399.78,308.5) .. controls (398.77,308.5) and (397.94,307.68) .. (397.94,306.66) -- cycle ;
\draw   (380.22,376.84) .. controls (381.87,377.07) and (383.53,378.21) .. (383.54,380.51) .. controls (383.55,385.1) and (376.95,385.12) .. (376.94,382.37) .. controls (376.93,379.61) and (383.53,379.59) .. (383.55,384.18) .. controls (383.57,388.78) and (376.96,388.8) .. (376.95,386.04) .. controls (376.94,383.29) and (383.55,383.26) .. (383.56,387.86) .. controls (383.58,392.45) and (376.97,392.47) .. (376.96,389.72) .. controls (376.95,386.96) and (383.56,386.94) .. (383.58,391.53) .. controls (383.59,396.13) and (376.99,396.15) .. (376.98,393.39) .. controls (376.97,390.64) and (383.57,390.61) .. (383.59,395.21) .. controls (383.61,399.8) and (377,399.82) .. (376.99,397.07) .. controls (376.98,394.31) and (383.59,394.29) .. (383.6,398.88) .. controls (383.62,403.48) and (377.01,403.5) .. (377,400.74) .. controls (376.99,397.99) and (383.6,397.96) .. (383.62,402.56) .. controls (383.62,402.89) and (383.58,403.19) .. (383.52,403.48) ;
\draw  [fill={rgb, 255:red, 0; green, 0; blue, 0 }  ,fill opacity=1 ] (380.72,404.25) .. controls (380.72,403.24) and (381.54,402.41) .. (382.56,402.41) .. controls (383.57,402.41) and (384.4,403.24) .. (384.4,404.25) .. controls (384.4,405.27) and (383.57,406.09) .. (382.56,406.09) .. controls (381.54,406.09) and (380.72,405.27) .. (380.72,404.25) -- cycle ;
\draw   (447.27,376.75) .. controls (445.46,376.99) and (443.65,378.14) .. (443.65,380.44) .. controls (443.67,385.03) and (450.93,385.01) .. (450.92,382.25) .. controls (450.91,379.49) and (443.65,379.52) .. (443.67,384.11) .. controls (443.68,388.71) and (450.94,388.68) .. (450.93,385.93) .. controls (450.92,383.17) and (443.66,383.2) .. (443.68,387.79) .. controls (443.7,392.38) and (450.95,392.36) .. (450.94,389.6) .. controls (450.93,386.84) and (443.68,386.87) .. (443.69,391.46) .. controls (443.71,396.06) and (450.97,396.03) .. (450.96,393.28) .. controls (450.95,390.52) and (443.69,390.55) .. (443.71,395.14) .. controls (443.72,399.73) and (450.98,399.71) .. (450.97,396.95) .. controls (450.96,394.19) and (443.7,394.22) .. (443.72,398.81) .. controls (443.74,403.41) and (451,403.38) .. (450.99,400.63) .. controls (450.98,397.87) and (443.72,397.9) .. (443.73,402.49) .. controls (443.74,403.4) and (444.03,404.13) .. (444.48,404.69) ;
\draw  [fill={rgb, 255:red, 0; green, 0; blue, 0 }  ,fill opacity=1 ] (443.72,405.25) .. controls (443.72,404.24) and (444.54,403.41) .. (445.56,403.41) .. controls (446.57,403.41) and (447.4,404.24) .. (447.4,405.25) .. controls (447.4,406.27) and (446.57,407.09) .. (445.56,407.09) .. controls (444.54,407.09) and (443.72,406.27) .. (443.72,405.25) -- cycle ;
\draw   (348.79,123.77) .. controls (348.79,90.79) and (375.58,64.05) .. (408.62,64.05) .. controls (441.66,64.05) and (468.45,90.79) .. (468.45,123.77) .. controls (468.45,156.75) and (441.66,183.49) .. (408.62,183.49) .. controls (375.58,183.49) and (348.79,156.75) .. (348.79,123.77) -- cycle ;
\draw   (348.79,123.77) .. controls (348.79,100.43) and (375.58,81.51) .. (408.62,81.51) .. controls (441.66,81.51) and (468.45,100.43) .. (468.45,123.77) .. controls (468.45,147.11) and (441.66,166.03) .. (408.62,166.03) .. controls (375.58,166.03) and (348.79,147.11) .. (348.79,123.77) -- cycle ;
\draw  [line width=1.5]  (407.7,60.37) -- (415.06,64.51) -- (407.7,68.64) ;
\draw  [line width=1.5]  (407.7,77.83) -- (415.06,81.97) -- (407.7,86.1) ;
\draw  [line width=1.5]  (415.38,169.4) -- (407.71,165.87) -- (414.72,161.16) ;
\draw  [line width=1.5]  (416.3,187.78) -- (408.63,184.24) -- (415.64,179.54) ;
\draw  [fill={rgb, 255:red, 0; green, 0; blue, 0 }  ,fill opacity=1 ] (423.32,101.74) .. controls (423.32,100.73) and (424.14,99.91) .. (425.16,99.91) .. controls (426.17,99.91) and (427,100.73) .. (427,101.74) .. controls (427,102.76) and (426.17,103.58) .. (425.16,103.58) .. controls (424.14,103.58) and (423.32,102.76) .. (423.32,101.74) -- cycle ;
\draw    (349.54,125.53) -- (396.63,101.74) ;
\draw    (468.45,123.77) -- (424.24,101.74) ;
\draw  [fill={rgb, 255:red, 0; green, 0; blue, 0 }  ,fill opacity=1 ] (392.94,102.66) .. controls (392.94,101.65) and (393.77,100.83) .. (394.78,100.83) .. controls (395.8,100.83) and (396.63,101.65) .. (396.63,102.66) .. controls (396.63,103.68) and (395.8,104.5) .. (394.78,104.5) .. controls (393.77,104.5) and (392.94,103.68) .. (392.94,102.66) -- cycle ;
\draw    (349.54,125.53) -- (394.33,142.5) ;
\draw  [fill={rgb, 255:red, 0; green, 0; blue, 0 }  ,fill opacity=1 ] (392.49,142.5) .. controls (392.49,141.48) and (393.32,140.66) .. (394.33,140.66) .. controls (395.35,140.66) and (396.17,141.48) .. (396.17,142.5) .. controls (396.17,143.51) and (395.35,144.33) .. (394.33,144.33) .. controls (393.32,144.33) and (392.49,143.51) .. (392.49,142.5) -- cycle ;
\draw    (468.45,123.77) -- (425.33,142.33) ;
\draw  [fill={rgb, 255:red, 0; green, 0; blue, 0 }  ,fill opacity=1 ] (422.65,142.33) .. controls (422.65,141.32) and (423.48,140.5) .. (424.49,140.5) .. controls (425.51,140.5) and (426.33,141.32) .. (426.33,142.33) .. controls (426.33,143.35) and (425.51,144.17) .. (424.49,144.17) .. controls (423.48,144.17) and (422.65,143.35) .. (422.65,142.33) -- cycle ;
\draw   (516.79,324.77) .. controls (516.79,291.79) and (543.58,265.05) .. (576.62,265.05) .. controls (609.66,265.05) and (636.45,291.79) .. (636.45,324.77) .. controls (636.45,357.75) and (609.66,384.49) .. (576.62,384.49) .. controls (543.58,384.49) and (516.79,357.75) .. (516.79,324.77) -- cycle ;
\draw   (516.79,324.77) .. controls (516.79,301.43) and (543.58,282.51) .. (576.62,282.51) .. controls (609.66,282.51) and (636.45,301.43) .. (636.45,324.77) .. controls (636.45,348.11) and (609.66,367.03) .. (576.62,367.03) .. controls (543.58,367.03) and (516.79,348.11) .. (516.79,324.77) -- cycle ;
\draw  [line width=1.5]  (575.7,261.37) -- (583.06,265.51) -- (575.7,269.64) ;
\draw  [line width=1.5]  (575.7,278.83) -- (583.06,282.97) -- (575.7,287.1) ;
\draw  [line width=1.5]  (582.46,351.11) -- (574.79,347.57) -- (581.8,342.87) ;
\draw  [line width=1.5]  (583.38,370.4) -- (575.71,366.87) -- (582.72,362.16) ;
\draw  [line width=1.5]  (584.3,388.78) -- (576.63,385.24) -- (583.64,380.54) ;
\draw  [draw opacity=0] (635.37,327.53) .. controls (632.15,338.89) and (607.26,348.02) .. (576.92,348.42) .. controls (544.56,348.85) and (518.14,339.2) .. (517.53,326.79) -- (576.62,325.69) -- cycle ; \draw   (635.37,327.53) .. controls (632.15,338.89) and (607.26,348.02) .. (576.92,348.42) .. controls (544.56,348.85) and (518.14,339.2) .. (517.53,326.79) ;  
\draw  [fill={rgb, 255:red, 0; green, 0; blue, 0 }  ,fill opacity=1 ] (591.32,302.74) .. controls (591.32,301.73) and (592.14,300.91) .. (593.16,300.91) .. controls (594.17,300.91) and (595,301.73) .. (595,302.74) .. controls (595,303.76) and (594.17,304.58) .. (593.16,304.58) .. controls (592.14,304.58) and (591.32,303.76) .. (591.32,302.74) -- cycle ;
\draw    (517.54,326.53) -- (564.63,302.74) ;
\draw    (635.37,327.52) -- (592.24,302.74) ;
\draw  [fill={rgb, 255:red, 0; green, 0; blue, 0 }  ,fill opacity=1 ] (560.94,303.66) .. controls (560.94,302.65) and (561.77,301.83) .. (562.78,301.83) .. controls (563.8,301.83) and (564.63,302.65) .. (564.63,303.66) .. controls (564.63,304.68) and (563.8,305.5) .. (562.78,305.5) .. controls (561.77,305.5) and (560.94,304.68) .. (560.94,303.66) -- cycle ;
\draw   (543.22,373.84) .. controls (544.87,374.07) and (546.53,375.21) .. (546.54,377.51) .. controls (546.55,382.1) and (539.95,382.12) .. (539.94,379.37) .. controls (539.93,376.61) and (546.53,376.59) .. (546.55,381.18) .. controls (546.57,385.78) and (539.96,385.8) .. (539.95,383.04) .. controls (539.94,380.29) and (546.55,380.26) .. (546.56,384.86) .. controls (546.58,389.45) and (539.97,389.47) .. (539.96,386.72) .. controls (539.95,383.96) and (546.56,383.94) .. (546.58,388.53) .. controls (546.59,393.13) and (539.99,393.15) .. (539.98,390.39) .. controls (539.97,387.64) and (546.57,387.61) .. (546.59,392.21) .. controls (546.61,396.8) and (540,396.82) .. (539.99,394.07) .. controls (539.98,391.31) and (546.59,391.29) .. (546.6,395.88) .. controls (546.62,400.48) and (540.01,400.5) .. (540,397.74) .. controls (539.99,394.99) and (546.6,394.96) .. (546.62,399.56) .. controls (546.62,399.89) and (546.58,400.19) .. (546.52,400.48) ;
\draw  [fill={rgb, 255:red, 0; green, 0; blue, 0 }  ,fill opacity=1 ] (543.72,401.25) .. controls (543.72,400.24) and (544.54,399.41) .. (545.56,399.41) .. controls (546.57,399.41) and (547.4,400.24) .. (547.4,401.25) .. controls (547.4,402.27) and (546.57,403.09) .. (545.56,403.09) .. controls (544.54,403.09) and (543.72,402.27) .. (543.72,401.25) -- cycle ;
\draw   (610.27,373.75) .. controls (608.46,373.99) and (606.65,375.14) .. (606.65,377.44) .. controls (606.67,382.03) and (613.93,382.01) .. (613.92,379.25) .. controls (613.91,376.49) and (606.65,376.52) .. (606.67,381.11) .. controls (606.68,385.71) and (613.94,385.68) .. (613.93,382.93) .. controls (613.92,380.17) and (606.66,380.2) .. (606.68,384.79) .. controls (606.7,389.38) and (613.95,389.36) .. (613.94,386.6) .. controls (613.93,383.84) and (606.68,383.87) .. (606.69,388.46) .. controls (606.71,393.06) and (613.97,393.03) .. (613.96,390.28) .. controls (613.95,387.52) and (606.69,387.55) .. (606.71,392.14) .. controls (606.72,396.73) and (613.98,396.71) .. (613.97,393.95) .. controls (613.96,391.19) and (606.7,391.22) .. (606.72,395.81) .. controls (606.74,400.41) and (614,400.38) .. (613.99,397.63) .. controls (613.98,394.87) and (606.72,394.9) .. (606.73,399.49) .. controls (606.74,400.4) and (607.03,401.13) .. (607.48,401.69) ;
\draw  [fill={rgb, 255:red, 0; green, 0; blue, 0 }  ,fill opacity=1 ] (606.72,402.25) .. controls (606.72,401.24) and (607.54,400.41) .. (608.56,400.41) .. controls (609.57,400.41) and (610.4,401.24) .. (610.4,402.25) .. controls (610.4,403.27) and (609.57,404.09) .. (608.56,404.09) .. controls (607.54,404.09) and (606.72,403.27) .. (606.72,402.25) -- cycle ;
\draw   (577.91,283.48) .. controls (579.22,283.82) and (580.54,285.2) .. (580.55,287.95) .. controls (580.57,293.47) and (575.33,293.48) .. (575.32,290.73) .. controls (575.31,287.97) and (580.55,287.95) .. (580.57,293.47) .. controls (580.59,298.98) and (575.35,299) .. (575.34,296.24) .. controls (575.33,293.48) and (580.57,293.47) .. (580.59,298.98) .. controls (580.59,300.19) and (580.34,301.14) .. (579.95,301.85) ;
\draw  [fill={rgb, 255:red, 0; green, 0; blue, 0 }  ,fill opacity=1 ] (577.19,303.71) .. controls (577.19,302.7) and (578.02,301.88) .. (579.03,301.88) .. controls (580.05,301.88) and (580.87,302.7) .. (580.87,303.71) .. controls (580.87,304.73) and (580.05,305.55) .. (579.03,305.55) .. controls (578.02,305.55) and (577.19,304.73) .. (577.19,303.71) -- cycle ;
\end{tikzpicture}
    \caption{Typical Feynman diagrams for the two-point correlation functions, where the thick lines represent light quarks, the spiral lines denote gluons, and the solid dots indicate the condensates. The diagrams include contributions from the perturbative term, quark condensates, gluon condensates, and mixed condensates.}
    \label{Feymann}
\end{figure}

Through the Källén-Lehmann spectral representation
\begin{equation}
    \rho(s) = \frac{1}{\pi} \Im \Pi(s),
\end{equation}
we can correspond the correlation functions given in Eqs.\eqref{1pLambda0}--\eqref{2psigma1} to the spectral density and derive the spectral density in the form of the operator product expansion (OPE). The spectral density, including vacuum condensates up to dimension 13, can generally be expressed as
    \begin{equation}
\begin{aligned}
    \rho^{\mathrm{OPE}}(s) =& \rho^{\mathrm{pert}}(s) 
+ \rho^{\langle \bar{q} q \rangle}(s) 
+ \rho^{\langle G^2 \rangle}(s) 
+ \rho^{\langle \bar{q} G q \rangle}(s) 
+ \rho^{\langle \bar{q} q \rangle^2}(s) 
+ \rho^{\langle G^3 \rangle}(s)+\rho^{\langle \bar{q} q \rangle \langle G^2 \rangle}(s)  \\
+& \rho^{\langle \bar{q} q \rangle \langle\bar{q}Gq\rangle}(s)+ \rho^{\langle G^4 \rangle}(s) + \rho^{\langle \bar{q} q \rangle^3}(s) + \rho^{\langle \bar{q} G q \rangle \langle G^2 \rangle}(s) 
+ \rho^{\langle \bar{q} q \rangle^2 \langle G^2 \rangle}(s)
+ \rho^{\langle \bar{q} G q \rangle^2}(s)\\
+& \rho^{\langle \bar{q} q \rangle^2 \langle\bar{q}Gq\rangle}(s)+\rho^{\langle \bar{q} q \rangle \langle G^4 \rangle}(s)
+ \rho^{\langle \bar{q} q \rangle^4}(s)+\rho^{\langle \bar{q} q \rangle\langle\bar{q}Gq\rangle \langle G^2 \rangle}(s)+\rho^{\langle \bar{q} q \rangle \langle\bar{q}Gq\rangle^2}(s)\\
+&\rho^{\langle \bar{q}G q \rangle \langle G^4 \rangle}(s).
\end{aligned}
\end{equation}
Subsequently, through the dispersion relation, the spectral density on the OPE side can be used to express the correlation function $\Pi_{X,J^{P}}^{\text{OPE}}(q^2)$ as 
\begin{equation}\label{OPE}
    \Pi_{X,J^{P}}^{\text{OPE}}(q^2) = \int_{s_{\text{min}}}^{\infty} \dd s \ \frac{\rho_{X,J^{P}}^{\text{OPE}}(s)}{s - q^2},
\end{equation}
where $X$ denotes the corresponding ground hadronic state and $J^{P}$ denotes its quantum number; $s_{\text{min}}$ represents the kinematic threshold, typically corresponding to the sum of the masses of all quarks involved in the hadronic interpolating current  \cite{Wan:2020oxt,Wan:2020fsk,Wan:2021vny}, i.e. $s_{\text{min}}=m_s^2$ for $p\bar\Lambda$ and $p\bar\Sigma$ molecular states. The analytical results of $\rho_{X,J^{P}}^{\text{OPE}}(s)$ are shown in the appendices.

\subsubsection{Phenomenological Side}
In the phenomenological side, the contributions from the ground state and excited states to the spectral density can be separated as
\begin{equation}
    \rho_{X,J^{P}}^{\text{Phen}}(s) = \lambda_{X,J^{P}}^2 \delta\left(s - m_{X,J^{P}}^2\right) + \theta\left(s - s_0\right) \rho_{X,J^{P}}(s),
\end{equation}
where $\lambda_{X,J^{P}}$ and $m_{X,J^{P}}$ represent the decay constant and mass of the ground state, respectively; $s_0$ represents the threshold of the continuum spectrum, above which the spectral density should include contributions from higher excited states and the continuum spectrum. Consequently, the phenomenological correlation function can be expressed through the dispersion relation as
\begin{equation}\label{phen}
    \Pi_{X,J^{P}}^{\text{Phen}}\left(q^2\right) = \frac{\lambda_{X,J^{P}}^2}{m_{X,J^{P}}^2 - q^2} + \int_{s_0}^{+\infty} \dd s \ \frac{\rho_{X,J^{P}}(s)}{s - q^2},
\end{equation}
where the first term corresponds to the contribution from the ground state, while the second term arises from higher excited states and the continuum spectrum.

\subsection{Hadronic Mass and Decay Constants}

According to the principle of quark-hadron duality , the correlation functions from both the OPE and the phenomenological sides should be consistent. Above the continuum threshold $s_0$, the spectral density satisfies $\rho_{X,J^{P}}(s) \approx \rho_{X,J^{P}}^{\text{OPE}}(s)$. Based on this principle, we combine the two sides of the correlation function Eqs.\eqref{OPE} and \eqref{phen} and apply the Borel transformation to both sides, which suppresses the contributions from higher excited states and the continuum, yielding
\begin{equation}\label{SR}
    \lambda_{X,J^{P}}^2 \mathrm{e}^{-m_{X,J^{P}}^2 / M_B^2} = \int_{s_{\min}}^{s_0} \dd s \ \rho_{X,J^{P}}^{\text{OPE}}(s) \mathrm{e}^{-s / M_B^2}.
\end{equation}
From the sum rule Eq.\eqref{SR}, the mass of the ground-state hadron $X$ can be determined as
\begin{equation}\label{mass}
    m_{X,J^{P}}\left(s_0, M_B^2\right) = \sqrt{-\frac{L_{X,J^{P},1}\left(s_0, M_B^2\right)}{L_{X,J^{P},0}\left(s_0, M_B^2\right)}},
\end{equation}
where
\begin{equation}
    \begin{aligned}\label{moment}
        &L_{X,J^{P},0}\left(s_0, M_B^2\right) = \int_{s_{\min}}^{s_0} \dd s \ \rho^{\text{OPE}}(s) \mathrm{e}^{-s / M_B^2} + \Pi^{\text{sum}}(M_B^2), \\
        &L_{X,J^{P},1}\left(s_0, M_B^2\right) = \frac{\partial}{\partial M_B^{-2}} L_{X,J^{P},0}\left(s_0, M_B^2\right).
    \end{aligned}
\end{equation}
Here, $\Pi^{\text{sum}}(M_B^2)$ represents the part of the correlation function that has no imaginary part but provides non-trivial contributions after the Borel transformation, which are proportional to the masses of $u$ and $d$ quarks. Since we have taken the limit $m_u = m_d \to 0$, this contribution vanishes and does not appear in our calculations. The decay constants can also be evaluated from the sum rule Eq.\eqref{SR} after the masses are determined, i.e.
\begin{equation}
	 \lambda_{X,J^{P}}\left(s_0, M_B^2\right) =\sqrt{\mathrm{e}^{m_{X,J^{P}}^2\left(s_0, M_B^2\right) / M_B^2}L_{X,J^{P},0}\left(s_0, M_B^2\right) }.
\end{equation}

\section{Numerical Analysis\label{na}}
In the numerical calculations of QCDSR, we adopt the following input parameters \cite{Colangelo:2000dp,Tang:2019nwv,Wan:2021vny}, where $q$ represents the $u,d$ quarks:
\begin{equation}
\begin{array}{ll}
    \langle \bar{q}q \rangle = -(0.24 \pm 0.01)^3 \ \text{GeV}^3, & 
    \langle \bar{s}s \rangle = (1.15 \pm 0.12)\langle \bar{q}q \rangle, \\
    \langle g_s^2 G^2 \rangle = (0.88 \pm 0.25) \ \text{GeV}^4, &
    \langle g_s^3 G^3 \rangle = (0.045 \pm 0.013) \ \text{GeV}^6, \\
    \langle \bar{q}g_s \sigma \cdot G q \rangle = m_0^2 \langle \bar{q}q \rangle, &
    \langle \bar{s}g_s \sigma \cdot G s \rangle = m_0^2 \langle \bar{s}s \rangle, \\
     m_0^2 = (0.8 \pm 0.1) \ \text{GeV}^2, &
    m_s = (95 \pm 5) \ \text{MeV}.
\end{array}
\end{equation}

In the process of establishing QCDSR, we also introduce two additional parameters, namely the continuum threshold parameter \( s_0 \) and the Borel parameter \( M_B^2 \). Theoretically, the mass \( m_X \)  and the decay constant $\lambda_X$ should not depend on these two parameters. Therefore, it is necessary to identify a range of \( s_0 \) and \( M_B^2 \) in which the variation of both the \( m_X \) and the $\lambda_{X}$ remains relatively stable. This range of parameters is the so-called Borel window. To determine an appropriate Borel window, we need to introduce two criteria. First, since we are calculating the mass of the ground-state hadron, the pole contribution should dominate.  According to Eq.\eqref{moment}, the contribution to the spectral density from \( s \gg M_B^2 \) is significantly suppressed \cite{Chen:2014vha,Azizi:2019xla,Wang:2017sto}. Therefore, the fraction of pole contribution can be defined as 
\begin{equation}
R^\text{PC}_{X,J^{P}} = \frac{L_{X,J^{P},0}(s_0,M_B^2)}{L_{X,J^{P},0}(\infty,M_B^2)}.
\end{equation}
For hexaquark states, we assume that this value should be no less than \( 15\%\) \cite{Wan:2019ake,Wan:2021vny}.  Secondly, the OPE should converge, which means that the contribution from the highest-dimensional condensate should be as small as possible. In this work, we calculate the contributions to the spectral density up to the dimension 13 condensates. The ratio of the dimension 13 condensate contribution is defined as  
\begin{equation}
R^{\text{OPE}}_{X,J^{P}} =\left\vert \frac{L^{\langle O_{13}\rangle}_{X,J^{P},0}(s_0,M_B^2)}{L_{X,J^{P},0}(s_0,M_B^2)}\right\vert,
\end{equation}
where
\begin{equation}
    L^{\langle \mathcal{O}_{13}\rangle}_{X,J^{P},0}(s_0,M_B^2)=\int_{s_{\min}}^{s_0} \dd s \ \rho^{\langle \mathcal{O}_{13}\rangle}(s) \mathrm{e}^{-s / M_B^2} .
\end{equation}

To determine the range of \( s_0 \), we adopt the method proposed in Refs.\cite{Wan:2020oxt,Wan:2020fsk,Wan:2021vny,Qiao:2013dda,Tang:2016pcf}. First, since \( s_0 \) represents the threshold for the onset of the continuum spectrum, it should be slightly larger than the hadron mass \( m_X \), that is, \( \sqrt{s_0} \sim m_X + \delta \), where \( \delta \) is typically taken as \( 0.4 - 0.8 \, \mathrm{GeV} \) \cite{Colangelo:2000dp,Wan:2021vny,Finazzo:2011he}. Next, within this interval of \( \sqrt{s_0} \), we search for the range of \( M_B \) where the mass \( m_X \) remains nearly stable. Since the contribution to the spectral density of \( s \gg M_B^2 \) is significantly suppressed, the $s\sim M_B$ region is pole dominated. Therefore, \( M_B \) should characterize the energy scale of the hadron \( X \). The range of \( s_0 \) and \( M_B \) that satisfies the above requirements is the Borel window, which corresponds to the physical state of the hadron under our investigation. In practice, we allow \( \sqrt{s_0} \) to vary within a range of \( \pm 0.1  \mathrm{GeV} \). A stable physical state is identified when the variation of \( M_B^2 \) within the Borel window exceeds \( 0.5 \, \mathrm{GeV}^2 \), ensuring the stability of the results.

\begin{figure}[H]
	\vspace{-1cm}
    \centering
    \subfigure[]{\includegraphics[width=0.45\textwidth]{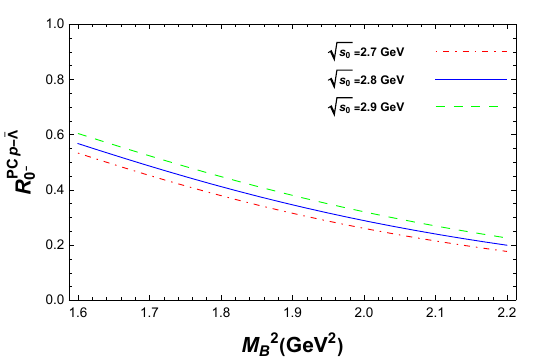}}
    \subfigure[]{\includegraphics[width=0.45\textwidth]{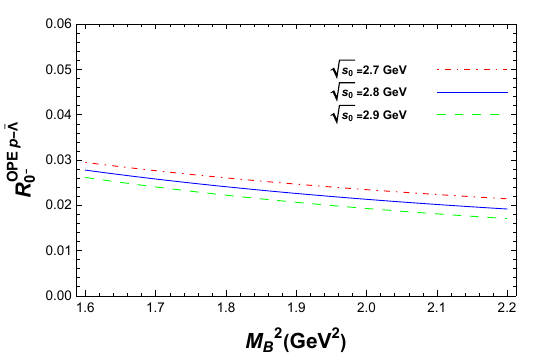}}
     \subfigure[]{\includegraphics[width=0.45\textwidth]{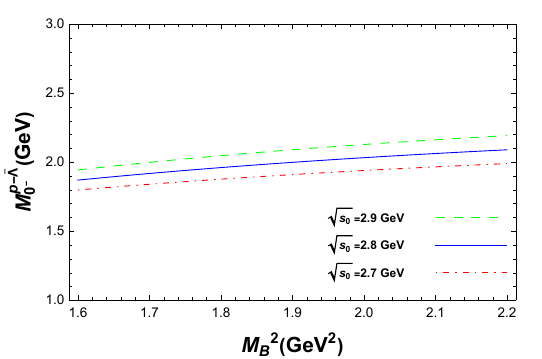}}
    \subfigure[]{\includegraphics[width=0.45\textwidth]{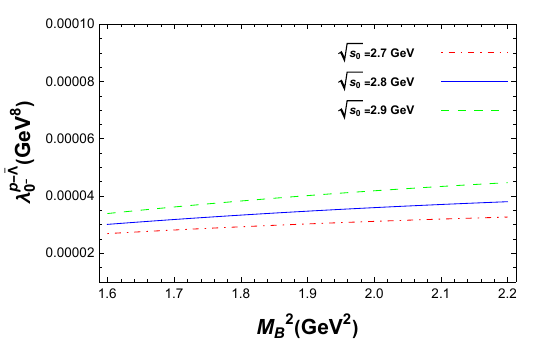}}
    \caption{The figures for the $0^{-}$ $p\bar\Lambda$ state. }
    \label{pLambda0-+fig}
\end{figure}

\begin{figure}[H]
	\vspace{-1.5cm}
    \centering
    \subfigure[]{\includegraphics[width=0.45\textwidth]{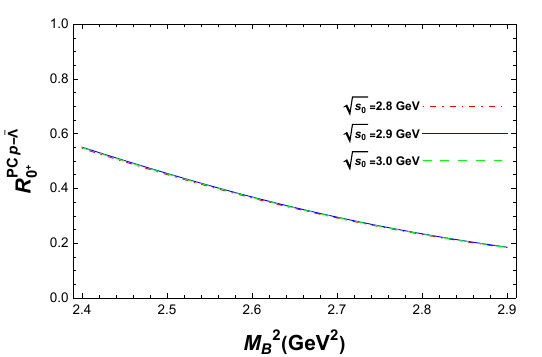}}
    \subfigure[]{\includegraphics[width=0.45\textwidth]{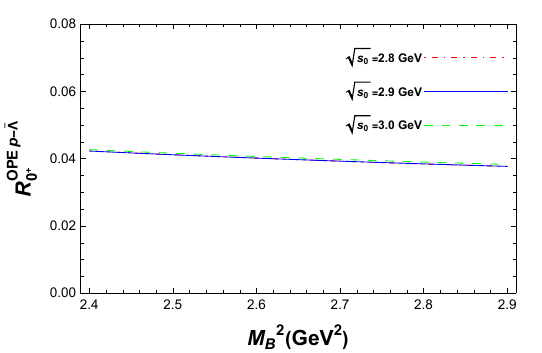}}
     \subfigure[]{\includegraphics[width=0.45\textwidth]{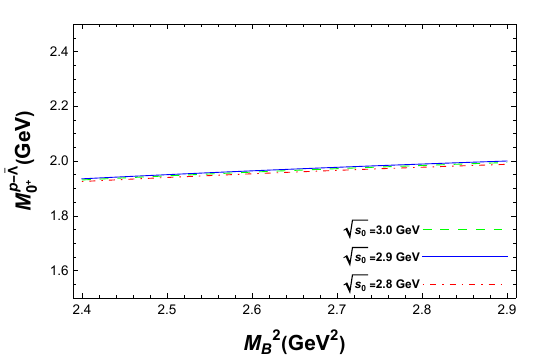}}
    \subfigure[]{\includegraphics[width=0.45\textwidth]{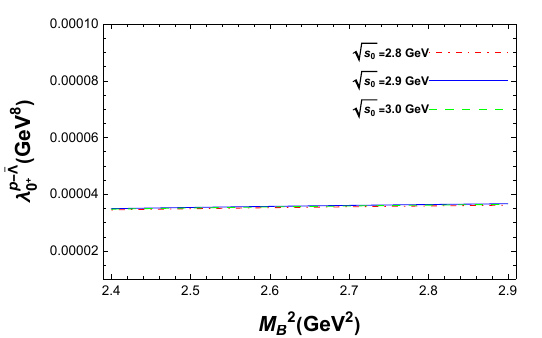}}
    \caption{The figures for the $0^{+}$ $p\bar\Lambda$ state. }
    \label{pLambda0++fig}
\end{figure}

\begin{figure}[H]
	\vspace{-1cm}
    \centering
    \subfigure[]{\includegraphics[width=0.45\textwidth]{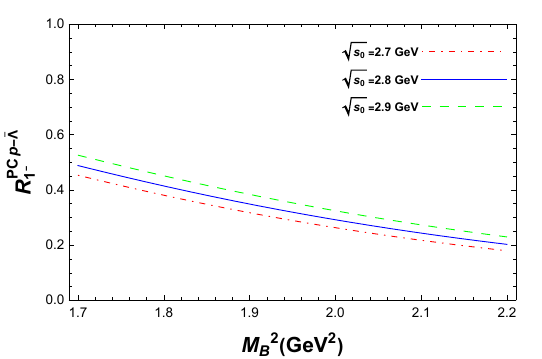}}
    \subfigure[]{\includegraphics[width=0.45\textwidth]{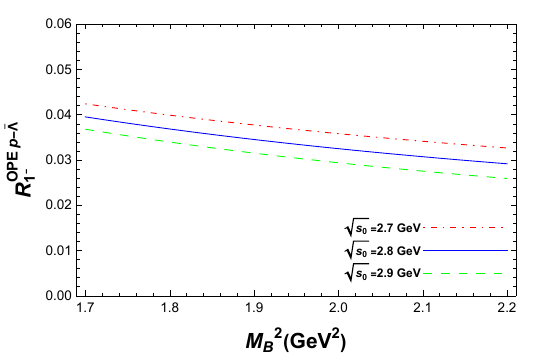}}
    \subfigure[]{\includegraphics[width=0.45\textwidth]{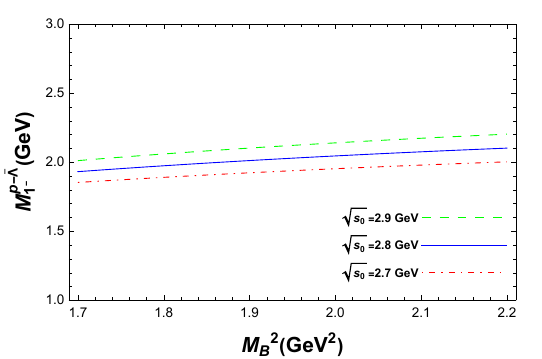}}
    \subfigure[]{\includegraphics[width=0.45\textwidth]{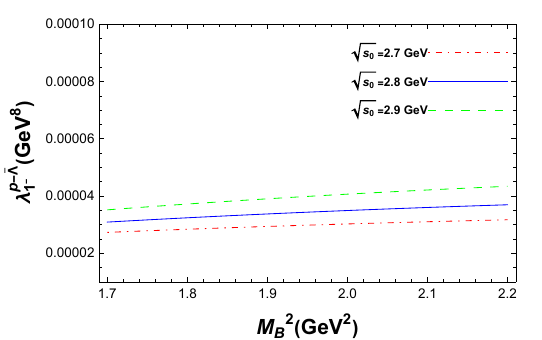}}
    \caption{The figures for the $1^{-}$ $p\bar\Lambda$ state. }
    \label{pLambda1--fig}
\end{figure}

\begin{figure}[H]
	\vspace{-1.5cm}
    \centering
    \subfigure[]{\includegraphics[width=0.45\textwidth]{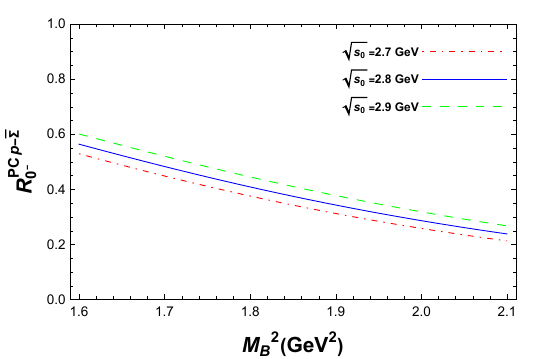}}
    \subfigure[]{\includegraphics[width=0.45\textwidth]{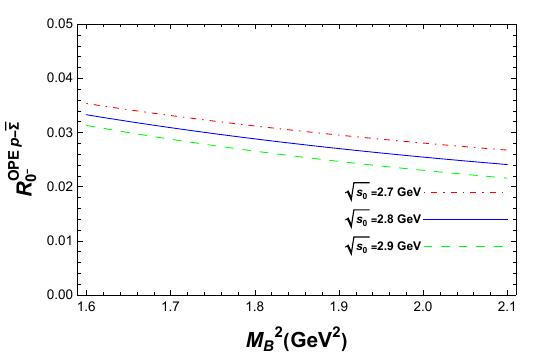}}
    \subfigure[]{\includegraphics[width=0.45\textwidth]{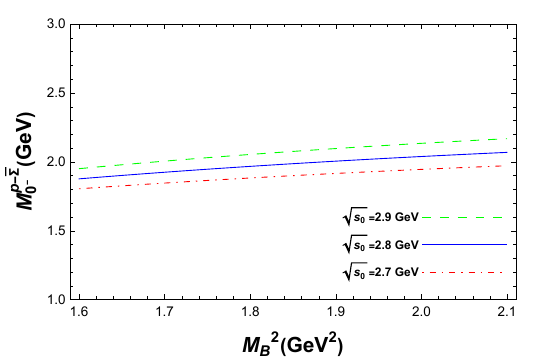}}
    \subfigure[]{\includegraphics[width=0.45\textwidth]{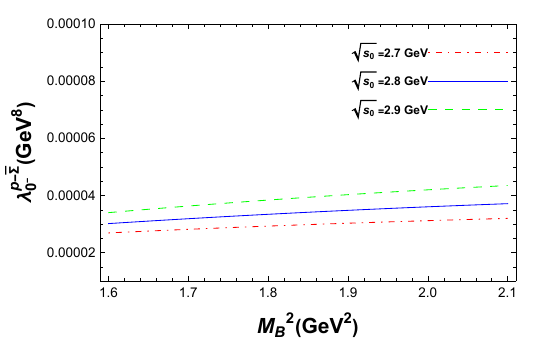}}
    \caption{The figures for the $0^{-}$ $p\bar\Sigma$ state.}
    \label{psigma0-+fig}
\end{figure}

\begin{figure}[H]
	\vspace{-1cm}
    \centering
    \subfigure[]{\includegraphics[width=0.45\textwidth]{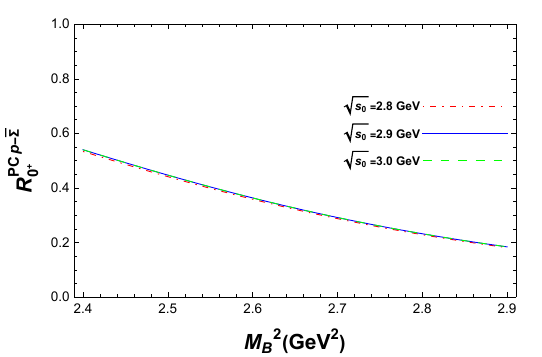}}
    \subfigure[]{\includegraphics[width=0.45\textwidth]{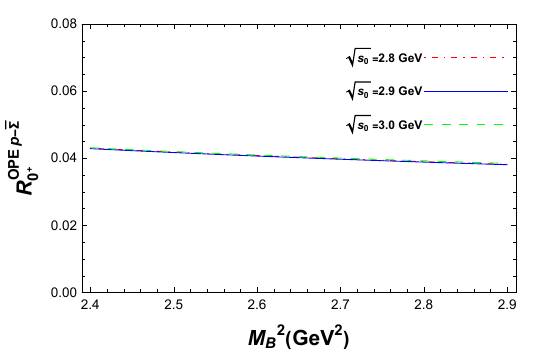}}
    \subfigure[]{\includegraphics[width=0.45\textwidth]{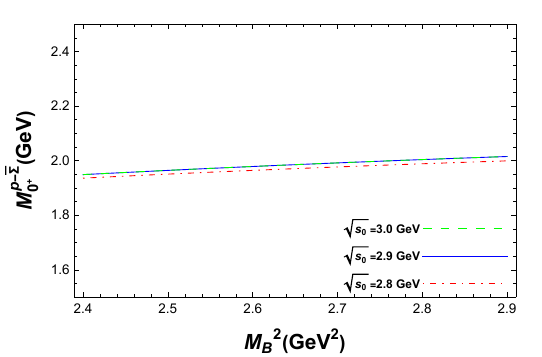}}
    \subfigure[]{\includegraphics[width=0.45\textwidth]{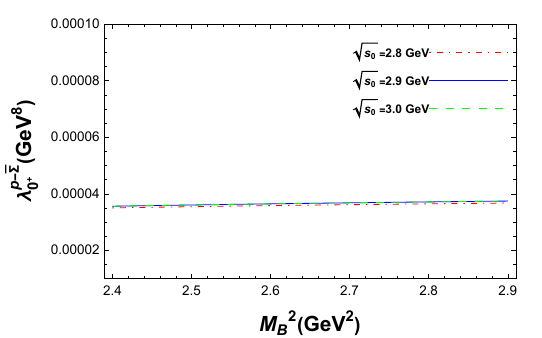}}
    \caption{The figures for the $0^{+}$ $p\bar\Sigma$ state. }
    \label{psigma0++fig}
\end{figure}

\begin{figure}[H]
	\vspace{-1.5cm}
    \centering
    \subfigure[]{\includegraphics[width=0.45\textwidth]{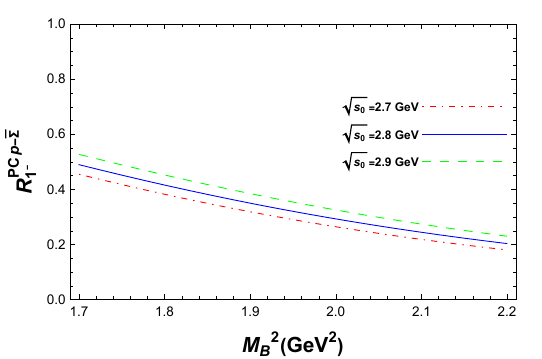}}
    \subfigure[]{\includegraphics[width=0.45\textwidth]{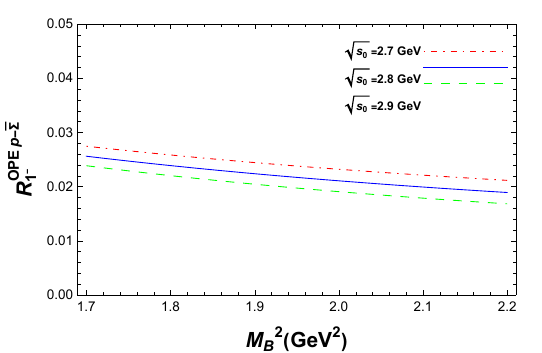}}
    \subfigure[]{\includegraphics[width=0.45\textwidth]{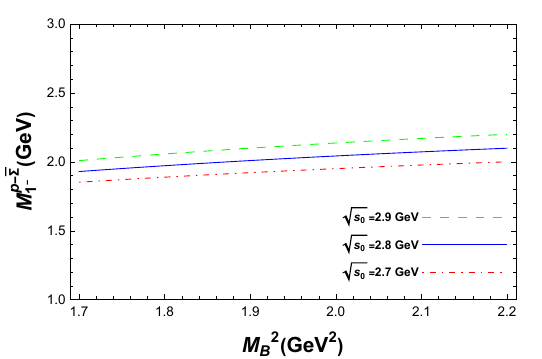}}
    \subfigure[]{\includegraphics[width=0.45\textwidth]{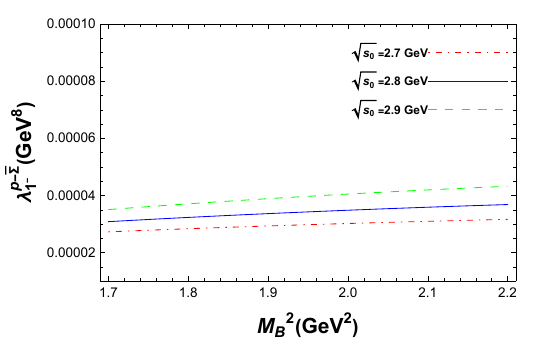}}
    \caption{The figures for the $1^{-}$ $p\bar\Sigma$ state. }
    \label{psigma1--fig}
\end{figure}

Based on the above discussion, we performed numerical calculations for the mass spectra of the ground states of the \( p\bar{\Lambda} \) and \( p\bar{\Sigma} \) molecular states. For both of the $p\bar{\Lambda}$ and $p\bar{\Sigma}$ states, we identified the Borel windows that satisfy the requirements for the $0^{+}$ states of the Type-1 current and the $0^{-}, 1^{-}$ states of the Type-2 current. For the other states, no suitable Borel window could be identified, regardless of the choice of \( s_0 \) and \( M_B^2 \). Therefore, we conclude that the corresponding current does not couple to such states, which is consistent with our previous work on the light baryonium \cite{Wan:2021vny}. The choices of Borel windows and the masses and decay constants of the states are shown in Table \ref{pLambdamass}. For the $0^{-}$ $p\bar\Lambda$ state, the ratios $R^{\text{PC}}_{p\bar\Lambda,0^{-}}$ and  $R^{\text{OPE}}_{p\bar\Lambda,0^{-}}$ are shown as functions of $M_B^2$ in Fig.\ref{pLambda0-+fig}(a), Fig.\ref{pLambda0-+fig}(b)with different values of $\sqrt{s_0}$, respectively. The dependence  between $m_{p\bar\Lambda,0^{-}}$ , $\lambda_{p\bar\Lambda,0^{-}}$ with $M_B^2$ is shown in Fig.\ref{pLambda0-+fig}(c), Fig.\ref{pLambda0-+fig}(d) with different values of $\sqrt{s_0}$, respectively. The same figures for $0^{+}$ and $1^{-}$ $p\bar\Lambda$ states are shown in Fig.\ref{pLambda0++fig}- Fig.\ref{pLambda1--fig} , respectively. Meanwhile, the same figures for $0^-$, $0^{+}$ and $1^{-}$ $p\bar\Sigma$ states are shown in Figs.\ref{psigma0-+fig}-\ref{psigma1--fig}, respectively.
\begin{table}[H]
	\small
	\begin{ruledtabular}
		\linespread{1.05}\selectfont
    \begin{tabular}{ccccccccc}
   Current &  $J^{P}$ &  State & $\sqrt{s_0}(\mathrm{GeV})$ & $M_B^2 (\mathrm{GeV}^2)$ & $M_X(\mathrm{GeV})$ & $\lambda_{X}(10^{-5}\mathrm{GeV}^8)$ & $R^{\mathrm{PC}}(\%)$ & $R^{\mathrm{OPE}}(\%)$\\
   \hline
   Type-1  
          & $0^{+}$ & $p\bar{\Lambda}$ & $2.9\pm 0.1$ & $2.4-2.9$ & $1.96\pm 0.03$ & $3.54\pm 0.10$& $18-55$&$3.8-4.2$ \\
          & &$p\bar{\Sigma}$&$2.9\pm 0.1$ & $2.4-2.9$ & $1.98\pm 0.04$& $3.62\pm 0.12$& $18-54$ & $3.8-4.3$ \\
    \hline
   Type-2  & $0^{-}$ & $p\bar{\Lambda}$ & $2.8\pm 0.1$ & $1.6-2.2$ & $2.00\pm 0.20$& $3.57\pm 0.89$& $18-60$&$1.7-2.9$ \\
   & &$p\bar{\Sigma}$& $2.8\pm 0.1$ & $1.6-2.1$ & $1.99\pm 0.18$& $3.52\pm0.83$ & $21-60$& $2.2-3.5$ \\
          & $1^{-}$ & $p\bar{\Lambda}$ & $2.8\pm 0.1$ & $1.7-2.2$ & $2.03\pm 0.17$  & $3.52\pm 0.81$ & $17-52$&$2.5-4.2$ \\
          & & $p\bar{\Sigma}$ & $2.8\pm 0.1$ & $1.7-2.2$ & $2.03\pm 0.17$&$3.52\pm 0.80$ & $18-53$&$1.6-2.7$ \\
\end{tabular}
\end{ruledtabular}
    \caption{The related numerical results of the 6 possible $p\bar\Lambda$ and $p\bar{\Sigma}$ molecular states}
    \label{pLambdamass}
\end{table}

From Table \ref{pLambdamass}, it can be observed that the mass results of the hexaquark states exhibit uncertainties. These uncertainties arise from the input quark masses, the vacuum condensates, and the choices of $\sqrt{s_0}$ and $M_B^2$. We also compared our results with those calculated through constituent quark models without annihilation interactions \cite{Huang:2011zq}, including the chiral quark model (ChQM) and the quark delocalization color-screening model (QDCSM) in Table.\ref{Comparison}. In this comparison, we adopted the mass data of \(p\), \(\Lambda\) and $\Sigma$ from the Particle Data Group (PDG) \cite{ParticleDataGroup:2024cfk}. It can be seen that the results for \(p\bar{\Lambda}\) obtained from both models are less than the corresponding dibaryon thresholds. However, for \(p\bar{\Sigma}\), the results from the chiral quark model exceed the corresponding dibaryon threshold, which is rather larger than the central values of our results.

\begin{table}[H]
    \centering
    \linespread{1.2}\selectfont
        \begin{tabular}{c|ccc|ccc|ccc}
        \hline\hline
        \diagbox{X}{$m_X(\mathrm{GeV})$}{Models} & & This work & & & QDCSM\cite{Huang:2011zq} & & & ChQM\cite{Huang:2011zq} & \\ 
        \hline
        $p\bar\Lambda$ $0^{-}$ & & 2.00 & & & \multirow{2}{*}{2.045} & & & \multirow{2}{*}{2.045} & \\
        $p\bar\Lambda$ $0^{+}$ & & 1.96 & & & & & & & \\
        \hline
        $p\bar\Lambda$ $1^{-}$ & & 2.03 & & & \multirow{2}{*}{2.048} & & & \multirow{2}{*}{2.047} &\\
        $p\bar\Lambda$ $1^{+}$ & & -- & & & & & & & \\
        \hline
        $p\bar\Sigma$ $0^{-}$ & & 1.99 & & & \multirow{2}{*}{2.132} & & & \multirow{2}{*}{2.165} &\\
        $p\bar\Sigma$ $0^{+}$ & & 1.98 & & & & & & & \\
        \hline
        $p\bar\Sigma$ $1^{-}$ & & 2.03 & & &  \multirow{2}{*}{2.152} & & &\multirow{2}{*}{2.189} &\\
        $p\bar\Sigma$ $1^{+}$ & & -- & & & & & & & \\
        \hline\hline
    \end{tabular}        
    \caption{The comparison between hexaquark state masses obtained in this work and the results from the constituent quark model.}
    \label{Comparison}
\end{table}

\section{Decay Modes Analyses \label{dm}}
In an experimental setting, the hexaquark states we have calculated may mix with other hadronic states due to coupling channel effects or mixing effects, such as excited states of the $K$ meson with the same quantum numbers and similar masses, which share the same strange numbers. These other hadronic states could mix into the background and influence detection. To address this, we can differentiate the hexaquark states we calculated from other states by measuring the decay modes of the hadronic states. In this section, we provide the possible decay modes of the six $p\bar\Lambda$ and $p\bar\Sigma$ molecular states studied earlier, with the hope that they can be observed on currently running experiments, such as BES\text{III}, BELLE\text{II}, and LHCb. 

Strong decay modes should be dominant, therefore we give only the strong decay modes. Since the central masses of these molecular states are all below their corresponding diquark thresholds, their direct strong decays into $p,\bar\Lambda$ or $p,\bar\Sigma$ may be forbidden. However, considering the uncertainty from calculation, the direct strong decays may be occur. The dominant decay modes are listed in Table.\ref{Decay}, while their mechanisms are shown in Fig.\ref{decaym}. 

In addition, the hexaquark states can also decay through the weak interaction. Both of $p\bar\Lambda$ and $p\bar\Sigma$ can decay into $p$ and $\bar p$. Since chiral limit is introduced in our calculation, the isospin triplet states of the $\Sigma$ baryon are degenerate. If isospin symmetry breaking is taken into account, the $p\bar\Sigma$ molecular states can also decay into $p$ and $\bar{n}$, while the isospin singlet $p\bar\Lambda$ states cannot. This difference could serve as a method to distinguish $p\bar\Lambda$ and $p\bar\Sigma$ in experiments. However, these weak decay processes are Cabibbo-suppressed, with rather small decay widths, making them potentially difficult to be observed experimentally.
\begin{table}[H]
    \begin{ruledtabular}
    	\linespread{1.3}\selectfont
       \begin{tabular}{cccc}
           $J^{P}$ & $0^{-}$ & $0^{+}$ & $1^{-}$  \\
           \hline
            Modes of $p\bar\Lambda$ ($p\bar\Sigma$)
            & $\pi\pi K$ & $\pi\pi K^{*}$ & $\pi\pi K^{*}$ \\
            & $\omega\omega K$ & $\omega\omega K^{*}$ & $\omega\omega K^{*}$ \\
       \end{tabular}
\end{ruledtabular}
    \caption{Typical decay modes of the $p\bar\Lambda$ and $p\bar\Sigma$ molecular states for each quantum number.}
    \label{Decay}
\end{table}
\begin{figure}[H]
	\centering

	\tikzset{every picture/.style={line width=0.75pt}} 
	
	\begin{tikzpicture}[x=0.75pt,y=0.75pt,yscale=-1,xscale=1]
		
		\draw  [fill={rgb, 255:red, 0; green, 0; blue, 0 }  ,fill opacity=0.69 ] (202.86,70.78) .. controls (220.09,70.76) and (234.09,102.73) .. (234.13,142.2) .. controls (234.17,181.66) and (220.24,213.66) .. (203.01,213.68) .. controls (185.78,213.7) and (171.78,181.72) .. (171.74,142.26) .. controls (171.69,102.8) and (185.63,70.79) .. (202.86,70.78) -- cycle ;
		\draw    (202.85,70.77) -- (327.3,69.73) ;
		\draw    (327.3,69.73) -- (419.32,213.58) ;
		\draw    (226.57,190.04) -- (326.66,190.04) ;
		\draw    (326.66,190.04) -- (419.32,213.58) ;
		\draw   (327.6,190.04) .. controls (327.02,188.48) and (326.66,186.46) .. (326.66,183.93) .. controls (326.66,171.36) and (335.59,171.36) .. (335.59,177.95) .. controls (335.59,184.53) and (326.66,184.53) .. (326.66,171.96) .. controls (326.66,159.39) and (335.59,159.39) .. (335.59,165.98) .. controls (335.59,172.56) and (326.66,172.56) .. (326.66,159.99) .. controls (326.66,147.42) and (335.59,147.42) .. (335.59,154) .. controls (335.59,160.59) and (326.66,160.59) .. (326.66,148.02) .. controls (326.66,135.45) and (335.59,135.45) .. (335.59,142.03) .. controls (335.59,148.62) and (326.66,148.62) .. (326.66,136.05) .. controls (326.66,123.48) and (335.59,123.48) .. (335.59,130.06) .. controls (335.59,136.65) and (326.66,136.65) .. (326.66,124.08) .. controls (326.66,111.51) and (335.59,111.51) .. (335.59,118.09) .. controls (335.59,124.68) and (326.66,124.68) .. (326.66,112.11) .. controls (326.66,99.54) and (335.59,99.54) .. (335.59,106.12) .. controls (335.59,112.71) and (326.66,112.71) .. (326.66,100.14) .. controls (326.66,87.57) and (335.59,87.57) .. (335.59,94.15) .. controls (335.59,100.73) and (326.66,100.73) .. (326.66,88.17) .. controls (326.66,75.6) and (335.59,75.6) .. (335.59,82.18) .. controls (335.59,88.76) and (326.66,88.76) .. (326.66,76.19) .. controls (326.66,73.47) and (327.08,71.33) .. (327.74,69.73) ;
		\draw  [fill={rgb, 255:red, 139; green, 87; blue, 42 }  ,fill opacity=1 ] (413.9,213.58) .. controls (413.9,210.77) and (416.32,208.49) .. (419.32,208.49) .. controls (422.31,208.49) and (424.73,210.77) .. (424.73,213.58) .. controls (424.73,216.39) and (422.31,218.67) .. (419.32,218.67) .. controls (416.32,218.67) and (413.9,216.39) .. (413.9,213.58) -- cycle ;
		\draw  [line width=1.5]  (263.44,67.56) -- (268.13,70.03) -- (263.44,72.51) ;
		\draw  [line width=1.5]  (358,112.36) -- (358.16,117.41) -- (353.48,114.91) ;
		\draw  [line width=1.5]  (389.83,208.24) -- (385.91,204.77) -- (391.1,203.44) ;
		\draw  [line width=1.5]  (280.24,192.22) -- (275.55,189.74) -- (280.24,187.27) ;
		\draw    (220.2,82.3) -- (365.98,80.5) ;
		\draw    (365.98,80.5) -- (421.65,143.35) ;
		\draw    (220.84,201.41) -- (365.55,202.01) ;
		\draw    (365.55,202.01) -- (421.65,143.35) ;
		\draw  [fill={rgb, 255:red, 74; green, 144; blue, 226 }  ,fill opacity=1 ] (416.23,143.35) .. controls (416.23,140.54) and (418.66,138.26) .. (421.65,138.26) .. controls (424.65,138.26) and (427.07,140.54) .. (427.07,143.35) .. controls (427.07,146.16) and (424.65,148.44) .. (421.65,148.44) .. controls (418.66,148.44) and (416.23,146.16) .. (416.23,143.35) -- cycle ;
		\draw    (203.01,213.68) -- (313.89,213.1) ;
		\draw    (225.94,94.27) -- (313.25,93.99) ;
		\draw    (313.25,93.99) -- (421.87,68.13) ;
		\draw    (313.89,213.1) -- (421.87,68.13) ;
		\draw  [fill={rgb, 255:red, 139; green, 87; blue, 42 }  ,fill opacity=1 ] (416.45,68.13) .. controls (416.45,65.32) and (418.87,63.05) .. (421.87,63.05) .. controls (424.86,63.05) and (427.28,65.32) .. (427.28,68.13) .. controls (427.28,70.94) and (424.86,73.22) .. (421.87,73.22) .. controls (418.87,73.22) and (416.45,70.94) .. (416.45,68.13) -- cycle ;
		\draw  [line width=1.5]  (264.08,79.53) -- (268.77,82.01) -- (264.08,84.48) ;
		\draw  [line width=1.5]  (264.08,91.5) -- (268.77,93.98) -- (264.08,96.45) ;
		\draw  [line width=1.5]  (280.24,204.19) -- (275.55,201.71) -- (280.24,199.24) ;
		\draw  [line width=1.5]  (280.88,216.16) -- (276.19,213.69) -- (280.88,211.21) ;
		\draw  [line width=1.5]  (401.99,116.55) -- (402.15,121.6) -- (397.47,119.1) ;
		\draw  [line width=1.5]  (377.71,75.83) -- (382.84,77.38) -- (378.76,80.68) ;
		\draw  [line width=1.5]  (401.24,169.27) -- (395.96,170.21) -- (397.85,165.48) ;
		\draw  [line width=1.5]  (354.09,163.1) -- (349.18,165.18) -- (349.84,160.17) ;
		\draw   (365.21,200.81) .. controls (364.64,199.25) and (364.28,197.23) .. (364.28,194.71) .. controls (364.28,182.14) and (373.2,182.14) .. (373.2,188.72) .. controls (373.2,195.3) and (364.28,195.3) .. (364.28,182.74) .. controls (364.28,170.17) and (373.2,170.17) .. (373.2,176.75) .. controls (373.2,183.33) and (364.28,183.33) .. (364.28,170.76) .. controls (364.28,158.19) and (373.2,158.19) .. (373.2,164.78) .. controls (373.2,171.36) and (364.28,171.36) .. (364.28,158.79) .. controls (364.28,146.22) and (373.2,146.22) .. (373.2,152.81) .. controls (373.2,159.39) and (364.28,159.39) .. (364.28,146.82) .. controls (364.28,134.25) and (373.2,134.25) .. (373.2,140.84) .. controls (373.2,147.42) and (364.28,147.42) .. (364.28,134.85) .. controls (364.28,122.28) and (373.2,122.28) .. (373.2,128.87) .. controls (373.2,135.45) and (364.28,135.45) .. (364.28,122.88) .. controls (364.28,110.31) and (373.2,110.31) .. (373.2,116.9) .. controls (373.2,123.48) and (364.28,123.48) .. (364.28,110.91) .. controls (364.28,98.34) and (373.2,98.34) .. (373.2,104.92) .. controls (373.2,111.51) and (364.28,111.51) .. (364.28,98.94) .. controls (364.28,86.37) and (373.2,86.37) .. (373.2,92.95) .. controls (373.2,99.54) and (364.28,99.54) .. (364.28,86.97) .. controls (364.28,84.24) and (364.7,82.11) .. (365.35,80.5) ;
		\draw   (314.76,213.1) .. controls (314.22,211.54) and (313.89,209.53) .. (313.89,207) .. controls (313.89,194.43) and (322.18,194.43) .. (322.18,201.01) .. controls (322.18,207.6) and (313.89,207.6) .. (313.89,195.03) .. controls (313.89,182.46) and (322.18,182.46) .. (322.18,189.04) .. controls (322.18,195.63) and (313.89,195.63) .. (313.89,183.06) .. controls (313.89,170.49) and (322.18,170.49) .. (322.18,177.07) .. controls (322.18,183.66) and (313.89,183.66) .. (313.89,171.09) .. controls (313.89,158.52) and (322.18,158.52) .. (322.18,165.1) .. controls (322.18,171.68) and (313.89,171.68) .. (313.89,159.11) .. controls (313.89,146.55) and (322.18,146.55) .. (322.18,153.13) .. controls (322.18,159.71) and (313.89,159.71) .. (313.89,147.14) .. controls (313.89,134.57) and (322.18,134.57) .. (322.18,141.16) .. controls (322.18,147.74) and (313.89,147.74) .. (313.89,135.17) .. controls (313.89,122.6) and (322.18,122.6) .. (322.18,129.19) .. controls (322.18,135.77) and (313.89,135.77) .. (313.89,123.2) .. controls (313.89,110.63) and (322.18,110.63) .. (322.18,117.22) .. controls (322.18,123.8) and (313.89,123.8) .. (313.89,111.23) .. controls (313.89,98.66) and (322.18,98.66) .. (322.18,105.25) .. controls (322.18,111.83) and (313.89,111.83) .. (313.89,99.26) .. controls (313.89,97) and (314.16,95.15) .. (314.59,93.67) ;
		
		\draw (167.02,63.93) node [anchor=north west][inner sep=0.75pt]    {$p$};
		\draw (148.38,203.78) node [anchor=north west][inner sep=0.75pt]    {$\bar{\Lambda }(\bar{\Sigma })$};
		\draw (426.15,59.74) node [anchor=north west][inner sep=0.75pt]    {$\pi ( \omega )$};
		\draw (426.51,205.78) node [anchor=north west][inner sep=0.75pt]    {$\pi ( \omega )$};
		\draw (427.79,130.75) node [anchor=north west][inner sep=0.75pt]    {$K\left( K^{*}\right)$};

	\end{tikzpicture}
    \caption{The mechanism of the dominant strong decays of $p\bar{\Lambda}$ ($p\bar\Sigma$)}
    \label{decaym}
\end{figure}

\section{Discussion and Conclusions}\label{dc}

In this work, we investigated the ground state mass spectra of the hexaquark molecular states of $p\bar{\Lambda}$ and $p\bar{\Sigma}$ with quantum numbers $J^{P}=0^{-}, 0^{+}, 1^{-}, 1^{+}$ in the framework of QCDSR. Two independent octet baryonic currents were chosen to establish the QCDSR. The results revealed that only 6 molecular bound states with quantum numbers $J^{P}=0^{-}, 0^{+}, 1^{-}$ may exist, with masses below their corresponding dibaryon thresholds. Since the quantum number of $X(2085)$ was fixed in experiment as $J^P=1^{+}$, our result hence disfavors it to be purely in configurations of $p\bar{\Lambda}$ and $p\bar{\Sigma}$ molecular structure. On the other hand, the mass of $X(2075)$ lies in the vicinity of $1^{-}$ $p\bar{\Lambda}$ and $p\bar{\Sigma}$ molecular states, and hence it possibly possesses large components of $p\bar{\Lambda}$ and $p\bar{\Sigma}$. We also analyzed the possible decay modes of these six hexaquark states, leaving them to later experimental confirmation. The hadronic states $X(2075)$ and $X(2085)$ are located in the region around $2$ GeV, which is known to be highly complex in the experimental spectrum and may accommodate a variety of hadronic configurations. As such, it is possible that both $X(2075)$ and $X(2085)$ are not pure states, but rather mixtures of different hadronic components. For example, $X(2075)$ might be a mixture of tetraquark and hexaquark states. The mixing effects of the currents expressed in Eq.\eqref{j0-}--\eqref{j1+} with interpolating currents of other states who share the same quantum numbers and similar masses can be taken into account in future works to enhance the precision of our prediction.

\begin{acknowledgments}
We appreciate the enlightening discussion with Bing-Dong Wan, Liang Tang and Yu-Ming Wang. This work was supported in part by the National Key Research and Development Program of China under Contracts No. 2020YFA0406400, by the National Natural Science Foundation of China(NSFC) under the Grants 12475087 and 12235008. 
\end{acknowledgments}
 
\bibliography{references.bib}

\newpage
\begin{appendix}
In the appendix, the analytical results for the spectral densities are presented, corresponding to 16 different configurations, i.e., four quantum numbers for each of the two hexaquark states; $p\bar{\Lambda}$ and $p\bar{\Sigma}$ states with two different baryonic currents. In calculating the spectral densities, the \texttt{FeynCalc} package \cite{Shtabovenko:2020gxv,Shtabovenko:2016sxi,Mertig:1990an} was utilized to trace out the $\gamma$-matrices. 
 
We expand the spectral densities as
\begin{equation}
    \rho^{\text{OPE}}=\rho^{\text{pert}}+\sum_{n=3}^{13}\rho^{\langle\mathcal{O}_n\rangle},
\end{equation}
where $n$ denotes for the dimension of the vacuum condensates.

    \section{The Spectral Densities of $p\bar\Lambda$ Hexaquark States}

    \subsection{$0^{-}$ $p\bar\Lambda$ Hexaquark States}
     \subsubsection{Type-1 Current}
     \vspace{-1cm}
     \begin{eqnarray}
         \rho^{\text{pert}}&=&\frac{s^7}{92484403200 \pi ^{10}},\\
          \rho^{\langle\mathcal{O}_3\rangle}&=&-\frac{m_s s^5 (2 \langle \bar{q}q\rangle-\langle \bar{s}s\rangle)}{58982400 \pi ^8},\\
          \rho^{\langle\mathcal{O}_4\rangle}&=&\frac{\langle GG \rangle s^5}{943718400 \pi ^{10}},\\
           \rho^{\langle\mathcal{O}_5\rangle}&=&\frac{m_s s^4 (3 \langle \bar{q}Gq\rangle-\langle \bar{s}Gs\rangle)}{11796480 \pi ^8},\\
           \rho^{\langle\mathcal{O}_6\rangle}&=&\frac{\langle \bar{q}q\rangle s^4 (\langle \bar{s}s\rangle-2 \langle \bar{q}q\rangle)}{737280 \pi ^6},\\
           \rho^{\langle\mathcal{O}_7\rangle}&=&-\frac{m_s \langle GG \rangle s^3 (2 \langle \bar{q}q\rangle-\langle \bar{s}s\rangle)}{2359296 \pi ^8},\\
           \rho^{\langle\mathcal{O}_8\rangle}&=&\frac{s^3 (4 \langle \bar{q}q\rangle \langle \bar{q}Gq\rangle-\langle \bar{q}q\rangle \langle \bar{s}Gs\rangle-\langle \bar{s}s\rangle \langle \bar{q}Gq\rangle)}{147456 \pi ^6}+\frac{\langle GG \rangle^2 s^3}{3623878656 \pi ^{10}},\\
           \rho^{\langle\mathcal{O}_9\rangle}&=&\frac{m_s \langle \bar{q}q \rangle^2 s^2 (2 \langle \bar{q}q \rangle-\langle \bar{s}s \rangle)}{2304 \pi ^4}+\frac{m_s \langle GG \rangle \langle \bar{q}Gq \rangle s^2}{393216 \pi ^8},\\
            \rho^{\langle\mathcal{O}_{10}\rangle}&=&\frac{\langle \bar{q}q\rangle \langle GG \rangle s^2 (\langle \bar{s}s\rangle-\langle \bar{q}q\rangle)}{73728 \pi ^6}+\frac{\langle \bar{q}Gq\rangle s^2 (\langle \bar{s}Gs\rangle-2 \langle \bar{q}Gq\rangle)}{49152 \pi ^6},\\
            \rho^{\langle\mathcal{O}_{11}\rangle}&=&\frac{m_s \langle \bar{q}q\rangle s (\langle \bar{q}q\rangle (\langle \bar{s}Gs\rangle-9 \langle \bar{q}Gq\rangle)+3 \langle \bar{s}s\rangle \langle \bar{s}Gs\rangle)}{2304 \pi ^4}-\frac{m_s \langle GG \rangle^2 s (2 \langle \bar{q}q\rangle-\langle \bar{s}s\rangle)}{75497472 \pi ^8},\\
        \rho^{\langle\mathcal{O}_{12}\rangle}&=&\frac{\langle \bar{q}q\rangle^3 s (\langle \bar{q}q\rangle-2 \langle \bar{s}s\rangle)}{288 \pi ^2}+\frac{\langle GG \rangle s (2 \langle \bar{q}q\rangle \langle \bar{q}Gq\rangle-\langle \bar{q}q\rangle \langle \bar{s}Gs\rangle-\langle \bar{s}s\rangle \langle \bar{q}Gq\rangle)}{49152 \pi ^6},\\
              \rho^{\langle\mathcal{O}_{13}\rangle}&=&\frac{m_s \langle \bar{q}Gq\rangle (18 \langle \bar{q}q\rangle \langle \bar{q}Gq\rangle-4 \langle \bar{q}q\rangle \langle \bar{s}Gs\rangle-3 \langle \bar{s}s\rangle \langle \bar{q}Gq\rangle)}{9216 \pi ^4} \nonumber\\
               &+&\frac{m_s \langle GG \rangle^2 (3 \langle \bar{q}Gq\rangle-\langle \bar{s}Gs\rangle)}{226492416 \pi ^8}.
     \end{eqnarray}
 \subsubsection{Type-2 Current}
   \vspace{-1cm}
    \begin{eqnarray}
         \rho^{\text{pert}}&=&\frac{s^7}{92484403200 \pi ^{10}},\\
          \rho^{\langle\mathcal{O}_3\rangle}&=&-\frac{m_s s^5 (2 \langle \bar{q}q\rangle-\langle \bar{s}s\rangle)}{58982400 \pi ^8},\\
          \rho^{\langle\mathcal{O}_4\rangle}&=&\frac{\langle GG \rangle s^5}{943718400 \pi ^{10}},\\
           \rho^{\langle\mathcal{O}_5\rangle}&=&\frac{m_s s^4 (3 \langle \bar{q}Gq\rangle-\langle \bar{s}Gs\rangle)}{11796480 \pi ^8},\\
           \rho^{\langle\mathcal{O}_6\rangle}&=&\frac{\langle \bar{q}q \rangle s^4 (2 \langle \bar{q}q \rangle+\langle \bar{s}s \rangle)}{737280 \pi ^6},\\
           \rho^{\langle\mathcal{O}_7\rangle}&=&-\frac{m_s \langle GG \rangle s^3 (2 \langle \bar{q}q\rangle-\langle \bar{s}s\rangle)}{2359296 \pi ^8},\\
           \rho^{\langle\mathcal{O}_8\rangle}&=&\frac{\langle GG \rangle^2 s^3}{3623878656 \pi ^{10}}-\frac{s^3 (\langle \bar{q}q \rangle (4 \langle \bar{q}Gq \rangle+\langle \bar{s}Gs \rangle)+\langle \bar{s}s \rangle \langle \bar{q}Gq \rangle)}{147456 \pi ^6},\\
           \rho^{\langle\mathcal{O}_9\rangle}&=&\frac{m_s \langle GG \rangle \langle \bar{q}Gq \rangle s^2}{393216 \pi ^8}-\frac{m_s \langle \bar{q}q \rangle^2 s^2 (2 \langle \bar{q}q \rangle-\langle \bar{s}s \rangle)}{2304 \pi ^4},\\
            \rho^{\langle\mathcal{O}_{10}\rangle}&=&\frac{\langle \bar{q}q \rangle \langle GG \rangle s^2 (\langle \bar{q}q \rangle+\langle \bar{s}s \rangle)}{73728 \pi ^6}+\frac{\langle \bar{q}Gq \rangle s^2 (2 \langle \bar{q}Gq \rangle+\langle \bar{s}Gs \rangle)}{49152 \pi ^6},\\
            \rho^{\langle\mathcal{O}_{11}\rangle}&=&\frac{m_s \langle \bar{q}q \rangle s (9 \langle \bar{q}q \rangle \langle \bar{q}Gq \rangle-\langle \bar{q}q \rangle \langle \bar{s}Gs \rangle-3 \langle \bar{s}s \rangle \langle \bar{s}Gs \rangle)}{2304 \pi ^4}\nonumber\\
            &-&\frac{m_s \langle GG \rangle^2 s (2 \langle \bar{q}q \rangle-\langle \bar{s}s \rangle)}{75497472 \pi ^8},\\
           \rho^{\langle\mathcal{O}_{12}\rangle}&=&\frac{\langle \bar{q}q \rangle^3 s (\langle \bar{q}q \rangle+2 \langle \bar{s}s \rangle)}{288 \pi ^2}-\frac{\langle GG \rangle s (\langle \bar{q}q \rangle (2 \langle \bar{q}Gq \rangle+\langle \bar{s}Gs \rangle)+\langle \bar{s}s \rangle \langle \bar{q}Gq \rangle)}{49152 \pi ^6},\\
              \rho^{\langle\mathcal{O}_{13}\rangle}&=&\frac{m_s \langle \bar{q}Gq \rangle (-18 \langle \bar{q}q \rangle \langle \bar{q}Gq \rangle+4 \langle \bar{q}q \rangle \langle \bar{s}Gs \rangle+3 \langle \bar{s}s \rangle \langle \bar{q}Gq \rangle)}{9216 \pi ^4} \nonumber\\
              &+&\frac{m_s \langle GG \rangle^2 (3 \langle \bar{q}Gq \rangle-\langle \bar{s}Gs \rangle)}{226492416 \pi ^8}.
     \end{eqnarray}

\subsection{$0^{+}$ $p\bar\Lambda$ Hexaquark States}
     \subsubsection{Type-1 Current}
     \vspace{-1cm}
     \begin{eqnarray}
         \rho^{\text{pert}}&=&\frac{s^7}{92484403200 \pi ^{10}},\\
          \rho^{\langle\mathcal{O}_3\rangle}&=&\frac{m_s s^5 (2 \langle \bar{q}q\rangle+\langle \bar{s}s\rangle)}{58982400 \pi ^8},\\
          \rho^{\langle\mathcal{O}_4\rangle}&=&\frac{\langle GG \rangle s^5}{943718400 \pi ^{10}},\\
           \rho^{\langle\mathcal{O}_5\rangle}&=&-\frac{m_s s^4 (3 \langle \bar{q}Gq\rangle+\langle \bar{s}Gs\rangle)}{11796480 \pi ^8},\\
           \rho^{\langle\mathcal{O}_6\rangle}&=&-\frac{\langle \bar{q}q\rangle s^4 (\langle \bar{s}s\rangle+2 \langle \bar{q}q\rangle)}{737280 \pi ^6},\\
           \rho^{\langle\mathcal{O}_7\rangle}&=&\frac{m_s \langle GG \rangle s^3 (2 \langle \bar{q}q\rangle+\langle \bar{s}s\rangle)}{2359296 \pi ^8},\\
           \rho^{\langle\mathcal{O}_8\rangle}&=&-\frac{s^3 (4 \langle \bar{q}q\rangle \langle \bar{q}Gq\rangle+\langle \bar{q}q\rangle \langle \bar{s}Gs\rangle+\langle \bar{s}s\rangle \langle \bar{q}Gq\rangle)}{147456 \pi ^6}+\frac{\langle GG \rangle^2 s^3}{3623878656 \pi ^{10}},\\
           \rho^{\langle\mathcal{O}_9\rangle}&=&-\frac{m_s \langle \bar{q}q \rangle^2 s^2 (2 \langle \bar{q}q \rangle+\langle \bar{s}s \rangle)}{2304 \pi ^4}-\frac{m_s \langle GG \rangle \langle \bar{q}Gq \rangle s^2}{393216 \pi ^8},\\
            \rho^{\langle\mathcal{O}_{10}\rangle}&=&-\frac{\langle \bar{q}q\rangle \langle GG \rangle s^2 (\langle \bar{s}s\rangle+\langle \bar{q}q\rangle)}{73728 \pi ^6}-\frac{\langle \bar{q}Gq\rangle s^2 (\langle \bar{s}Gs\rangle+2 \langle \bar{q}Gq\rangle)}{49152 \pi ^6},\\
            \rho^{\langle\mathcal{O}_{11}\rangle}&=&\frac{m_s \langle \bar{q}q\rangle s (\langle \bar{q}q\rangle (\langle \bar{s}Gs\rangle+9 \langle \bar{q}Gq\rangle)+3 \langle \bar{s}s\rangle \langle \bar{s}Gs\rangle)}{2304 \pi ^4} \nonumber\\
            &+&\frac{m_s \langle GG \rangle^2 s (2 \langle \bar{q}q\rangle+\langle \bar{s}s\rangle)}{75497472 \pi ^8},\\
           \rho^{\langle\mathcal{O}_{12}\rangle}&=&\frac{\langle \bar{q}q\rangle^3 s (\langle \bar{q}q\rangle+2 \langle \bar{s}s\rangle)}{288 \pi ^2}+\frac{\langle GG \rangle s (2 \langle \bar{q}q\rangle \langle \bar{q}Gq\rangle+\langle \bar{q}q\rangle \langle \bar{s}Gs\rangle+\langle \bar{s}s\rangle \langle \bar{q}Gq\rangle)}{49152 \pi ^6},\\
         \rho^{\langle\mathcal{O}_{13}\rangle}&=&-\frac{m_s \langle \bar{q}Gq\rangle (18 \langle \bar{q}q\rangle \langle \bar{q}Gq\rangle+4 \langle \bar{q}q\rangle \langle \bar{s}Gs\rangle+3 \langle \bar{s}s\rangle \langle \bar{q}Gq\rangle)}{9216 \pi ^4} \nonumber\\
         &-&\frac{m_s \langle GG \rangle^2 (3 \langle \bar{q}Gq\rangle+\langle \bar{s}Gs\rangle)}{226492416 \pi ^8}.
     \end{eqnarray}

   \subsubsection{Type-2 Current}
   \vspace{-1cm}
    \begin{eqnarray}
         \rho^{\text{pert}}&=&\frac{s^7}{92484403200 \pi ^{10}},\\
          \rho^{\langle\mathcal{O}_3\rangle}&=&\frac{m_s s^5 (2 \langle \bar{q}q\rangle+\langle \bar{s}s\rangle)}{58982400 \pi ^8},\\
          \rho^{\langle\mathcal{O}_4\rangle}&=&\frac{\langle GG \rangle s^5}{943718400 \pi ^{10}},\\
           \rho^{\langle\mathcal{O}_5\rangle}&=&-\frac{m_s s^4 (3 \langle\bar{q}Gq\rangle+\langle\bar{s}Gs\rangle)}{11796480 \pi ^8},\\
           \rho^{\langle\mathcal{O}_6\rangle}&=&-\frac{\langle\bar{q} q\rangle s^4 (\langle\bar{s} s\rangle-2 \langle\bar{q} q\rangle)}{737280 \pi ^6},\\
           \rho^{\langle\mathcal{O}_7\rangle}&=&\frac{m_s \langle GG \rangle s^3 (2 \langle\bar{q} q\rangle+\langle\bar{s} s\rangle)}{2359296 \pi ^8},\\
           \rho^{\langle\mathcal{O}_8\rangle}&=&-\frac{s^3 (4 \langle\bar{q} q\rangle \langle\bar{q}Gq\rangle-\langle\bar{q} q\rangle \langle\bar{s}Gs\rangle-\langle\bar{s} s\rangle \langle\bar{q}Gq\rangle)}{147456 \pi ^6}+\frac{\langle GG \rangle^2 s^3}{3623878656 \pi ^{10}},\\
           \rho^{\langle\mathcal{O}_9\rangle}&=&-\frac{m_s \langle GG \rangle \langle\bar{q}Gq\rangle s^2}{393216 \pi ^8}+\frac{m_s \langle\bar{q} q\rangle^2 s^2 (2 \langle\bar{q} q\rangle+\langle\bar{s} s\rangle)}{2304 \pi ^4},\\
            \rho^{\langle\mathcal{O}_{10}\rangle}&=&-\frac{\langle\bar{q} q\rangle \langle GG \rangle s^2 (\langle\bar{s} s\rangle-\langle\bar{q} q\rangle)}{73728 \pi ^6}-\frac{\langle\bar{q}Gq\rangle s^2 (\langle\bar{s}Gs\rangle-2 \langle\bar{q}Gq\rangle)}{49152 \pi ^6},\\
            \rho^{\langle\mathcal{O}_{11}\rangle}&=&-\frac{m_s \langle\bar{q} q\rangle s (\langle\bar{q} q\rangle (9 \langle\bar{q}Gq\rangle+\langle\bar{s}Gs\rangle)+3 \langle\bar{s} s\rangle \langle\bar{s}Gs\rangle)}{2304 \pi ^4}\nonumber\\
            &+&\frac{m_s \langle GG \rangle^2 s (2 \langle\bar{q} q\rangle+\langle\bar{s} s\rangle)}{75497472 \pi ^8},\\
           \rho^{\langle\mathcal{O}_{12}\rangle}&=&-\frac{\langle GG \rangle s (2 \langle\bar{q} q\rangle \langle\bar{q}Gq\rangle-\langle\bar{q} q\rangle \langle\bar{s}Gs\rangle-\langle\bar{s} s\rangle \langle\bar{q}Gq\rangle)}{49152 \pi ^6}+\frac{\langle\bar{q} q\rangle^3 s (\langle\bar{q} q\rangle-2 \langle\bar{s} s\rangle)}{288 \pi ^2},\\
              \rho^{\langle\mathcal{O}_{13}\rangle}&=&-\frac{m_s \langle GG \rangle^2 (3 \langle\bar{q}Gq\rangle+\langle\bar{s}Gs\rangle)}{226492416 \pi ^8}\nonumber\\
              &+&\frac{m_s \langle\bar{q}Gq\rangle (18 \langle\bar{q} q\rangle \langle\bar{q}Gq\rangle+4 \langle\bar{q} q\rangle \langle\bar{s}Gs\rangle+3 \langle\bar{s} s\rangle \langle\bar{q}Gq\rangle)}{9216 \pi ^4}.
     \end{eqnarray}

     \subsection{$1^{-}$ $p\bar\Lambda$ Hexaquark States}
     \subsubsection{Type-1 Current}
     \vspace{-1cm}
     \begin{eqnarray}
         \rho^{\text{pert}}&=&\frac{s^7}{104044953600 \pi ^{10}},\\
          \rho^{\langle\mathcal{O}_3\rangle}&=&-\frac{m_s s^5 (7 \langle \bar{q}q \rangle-3 \langle \bar{s}s \rangle)}{206438400 \pi ^8},\\
          \rho^{\langle\mathcal{O}_4\rangle}&=&\frac{\langle GG \rangle s^5}{1101004800 \pi ^{10}},\\
           \rho^{\langle\mathcal{O}_5\rangle}&=&\frac{m_s s^4 (18 \langle \bar{q}G q \rangle-5 \langle \bar{s}G s \rangle)}{70778880 \pi ^8},\\
           \rho^{\langle\mathcal{O}_6\rangle}&=&-\frac{\langle \bar{q}q \rangle s^4 (5 \langle \bar{q}q \rangle-3 \langle \bar{s}s \rangle)}{2211840 \pi ^6},\\
           \rho^{\langle\mathcal{O}_7\rangle}&=&-\frac{m_s \langle GG \rangle s^3 (5 \langle \bar{q}q \rangle-2 \langle \bar{s}s \rangle)}{5898240 \pi ^8},\\
           \rho^{\langle\mathcal{O}_8\rangle}&=&-\frac{s^3 (-16 \langle \bar{q}q \rangle \langle \bar{q}G q \rangle+5 \langle \bar{q}q \rangle \langle \bar{s}G s \rangle+5 \langle \bar{s}s \rangle \langle \bar{q}G q \rangle)}{737280 \pi ^6}+\frac{\langle GG \rangle^2 s^3}{4529848320 \pi ^{10}},\\
           \rho^{\langle\mathcal{O}_9\rangle}&=&\frac{m_s \langle \bar{q}q \rangle^2 s^2 (8 \langle \bar{q}q \rangle-3 \langle \bar{s}s \rangle)}{9216 \pi ^4}+\frac{m_s \langle GG \rangle s^2 (4 \langle \bar{q}G q \rangle-\langle \bar{s}G s \rangle)}{1572864 \pi ^8},\\
            \rho^{\langle\mathcal{O}_{10}\rangle}&=&-\frac{\langle \bar{q}q \rangle \langle GG \rangle s^2 (3 \langle \bar{q}q \rangle-4 \langle \bar{s}s \rangle)}{294912 \pi ^6}-\frac{\langle \bar{q}G q \rangle s^2 (3 \langle \bar{q}G q \rangle-2 \langle \bar{s}G s \rangle)}{98304 \pi ^6},\\
            \rho^{\langle\mathcal{O}_{11}\rangle}&=&-\frac{m_s \langle GG \rangle^2 s (3 \langle \bar{q}q \rangle-\langle \bar{s}s \rangle)}{113246208 \pi ^8}\nonumber\\
            &-&\frac{m_s \langle \bar{q}q \rangle s (27 \langle \bar{q}q \rangle \langle \bar{q}G q \rangle-2 \langle \bar{q}q \rangle \langle \bar{s}G s \rangle-6 \langle \bar{s}s \rangle \langle \bar{q}G q \rangle)}{6912 \pi ^4},\\
           \rho^{\langle\mathcal{O}_{12}\rangle}&=&-\frac{\langle GG \rangle s (-4 \langle \bar{q}q \rangle \langle \bar{q}G q \rangle+3 \langle \bar{q}q \rangle \langle \bar{s}G s \rangle+3 \langle \bar{s}s \rangle \langle \bar{q}G q \rangle)}{147456 \pi ^6}\\
           &+&\frac{\langle \bar{q}q \rangle^3 s (\langle \bar{q}q \rangle-3 \langle \bar{s}s \rangle)}{432 \pi ^2},\\
              \rho^{\langle\mathcal{O}_{13}\rangle}&=&\frac{m_s \langle \bar{q}G q \rangle (-25 \expval{\bar{q}q} \langle \bar{q}G q \rangle+34 \expval{\bar{q}q} \langle \bar{s}G s \rangle+16 \langle \bar{s}s \rangle \langle \bar{q}G q \rangle)}{36864 \pi ^4}\nonumber\\
              &+&\frac{m_s \langle GG \rangle^2 (6 \langle \bar{q}G q \rangle-\langle \bar{s}G s \rangle)}{452984832 \pi ^8}.
     \end{eqnarray}

   \subsubsection{Type-2 Current}
   \vspace{-1cm}
    \begin{eqnarray}
         \rho^{\text{pert}}&=&\frac{s^7}{104044953600 \pi ^{10}},\\
          \rho^{\langle\mathcal{O}_3\rangle}&=&-\frac{m_s s^5 (7 \langle \bar{q}q \rangle-3 \langle \bar{s}s \rangle)}{206438400 \pi ^8},\\
          \rho^{\langle\mathcal{O}_4\rangle}&=&\frac{\langle GG \rangle s^5}{1101004800 \pi ^{10}},\\
           \rho^{\langle\mathcal{O}_5\rangle}&=&\frac{m_s s^4 (18 \langle \bar{q}G q \rangle-5 \langle \bar{s}G s \rangle)}{70778880 \pi ^8},\\
            \rho^{\langle\mathcal{O}_6\rangle}&=&\frac{\langle \bar{q}q \rangle s^4 (5 \langle \bar{q}q \rangle+3 \langle \bar{s}s \rangle)}{2211840 \pi ^6},\\
           \rho^{\langle\mathcal{O}_7\rangle}&=&-\frac{m_s \langle GG \rangle s^3 (5 \langle \bar{q}q \rangle-2 \langle \bar{s}s \rangle)}{5898240 \pi ^8},\\
           \rho^{\langle\mathcal{O}_8\rangle}&=&-\frac{s^3 (16 \langle \bar{q}q \rangle \langle \bar{q}G q \rangle+5 \langle \bar{q}q \rangle \langle \bar{s}G s \rangle+5 \langle \bar{s}s \rangle \langle \bar{q}G q \rangle)}{737280 \pi ^6}+\frac{\langle GG \rangle^2 s^3}{4529848320 \pi ^{10}},\\
           \rho^{\langle\mathcal{O}_9\rangle}&=&-\frac{m_s \langle \bar{q}q \rangle^2 s^2 (8 \langle \bar{q}q \rangle-3 \langle \bar{s}s \rangle)}{9216 \pi ^4}+\frac{m_s \langle GG \rangle s^2 (4 \langle \bar{q}G q \rangle-\langle \bar{s}G s \rangle)}{1572864 \pi ^8},\\
            \rho^{\langle\mathcal{O}_{10}\rangle}&=&\frac{\langle \bar{q}q \rangle \langle GG \rangle s^2 (3 \langle \bar{q}q \rangle+4 \langle \bar{s}s \rangle)}{294912 \pi ^6}+\frac{\langle \bar{q}G q \rangle s^2 (3 \langle \bar{q}G q \rangle+2 \langle \bar{s}G s \rangle)}{98304 \pi ^6},\\
            \rho^{\langle\mathcal{O}_{11}\rangle}&=&-\frac{m_s \langle GG \rangle^2 s (3 \langle \bar{q}q \rangle-\langle \bar{s}s \rangle)}{113246208 \pi ^8}\nonumber\\
            &+&\frac{m_s \langle \bar{q}q \rangle s (27 \langle \bar{q}q \rangle \langle \bar{q}G q \rangle-2 \langle \bar{q}q \rangle \langle \bar{s}G s \rangle-6 \langle \bar{s}s \rangle \langle \bar{q}G q \rangle)}{6912 \pi ^4},\\
           \rho^{\langle\mathcal{O}_{12}\rangle}&=&-\frac{\langle GG \rangle s (4 \langle \bar{q}q \rangle \langle \bar{q}G q \rangle+3 \langle \bar{q}q \rangle \langle \bar{s}G s \rangle+3 \langle \bar{s}s \rangle \langle \bar{q}G q \rangle)}{147456 \pi ^6}\\
           &+&\frac{\langle \bar{q}q \rangle^3 s (\langle \bar{q}q \rangle+3 \langle \bar{s}s \rangle)}{432 \pi ^2},\\
              \rho^{\langle\mathcal{O}_{13}\rangle}&=&\frac{m_s \expval{GG}^2 (6 \expval{\bar q G q}-\expval{\bar s G s})}{452984832 \pi ^8}\nonumber\\
              &-&\frac{m_s \langle \bar{q}G q \rangle (11 \expval{\bar{q}q} \langle \bar{q}G q \rangle+14 \expval{\bar{q}q} \langle \bar{s}G s \rangle+8 \langle \bar{s}s \rangle \langle \bar{q}G q \rangle)}{36864 \pi ^4}.
     \end{eqnarray}

\subsection{$1^{+}$ $p\bar\Lambda$ Hexaquark States}
     \subsubsection{Type-1 Current}
     \vspace{-1cm}
     \begin{eqnarray}
         \rho^{\text{pert}}&=&\frac{s^7}{104044953600 \pi ^{10}},\\
          \rho^{\langle\mathcal{O}_3\rangle}&=&\frac{m_s s^5 (7 \langle \bar{q}q \rangle+3 \langle \bar{s}s \rangle)}{206438400 \pi ^8},\\
          \rho^{\langle\mathcal{O}_4\rangle}&=&\frac{\langle GG \rangle s^5}{1101004800 \pi ^{10}},\\
           \rho^{\langle\mathcal{O}_5\rangle}&=&-\frac{m_s s^4 (18 \langle \bar{q}G q \rangle-5 \langle \bar{s}G s \rangle)}{70778880 \pi ^8},\\
           \rho^{\langle\mathcal{O}_6\rangle}&=&-\frac{\langle \bar{q}q \rangle s^4 (5 \langle \bar{q}q \rangle+3 \langle \bar{s}s \rangle)}{2211840 \pi ^6},\\
           \rho^{\langle\mathcal{O}_7\rangle}&=&\frac{m_s \langle GG \rangle s^3 (5 \langle \bar{q}q \rangle+2 \langle \bar{s}s \rangle)}{5898240 \pi ^8},\\
           \rho^{\langle\mathcal{O}_8\rangle}&=&\frac{s^3 (16 \langle \bar{q}q \rangle \langle \bar{q}G q \rangle+5 \langle \bar{q}q \rangle \langle \bar{s}G s \rangle+5 \langle \bar{s}s \rangle \langle \bar{q}G q \rangle)}{737280 \pi ^6}+\frac{\langle GG \rangle^2 s^3}{4529848320 \pi ^{10}},\\
           \rho^{\langle\mathcal{O}_9\rangle}&=&-\frac{m_s s^2 \left(512 \pi ^4 \langle \bar{q}q \rangle^2 (8 \langle \bar{q}q \rangle+3 \langle \bar{s}s \rangle)+3 \langle GG \rangle (4 \langle \bar{q}G q \rangle+\langle \bar{s}G s \rangle)\right)}{4718592 \pi ^8},\\
            \rho^{\langle\mathcal{O}_{10}\rangle}&=&-\frac{s^2 \left(3 \langle \bar{q}q \rangle^2 \langle GG \rangle+4 \langle \bar{q}q \rangle \langle \bar{s}s \rangle \langle GG \rangle+9 \langle \bar{q}G q \rangle^2+6 \langle \bar{q}G q \rangle \langle \bar{s}G s \rangle\right)}{294912 \pi ^6},\\
            \rho^{\langle\mathcal{O}_{11}\rangle}&=&\frac{m_s \langle GG \rangle^2 s (3 \langle \bar{q}q \rangle+\langle \bar{s}s \rangle)}{113246208 \pi ^8}\nonumber\\
            &+&\frac{m_s \langle \bar{q}q \rangle s (27 \langle \bar{q}q \rangle \langle \bar{q}G q \rangle+2 \langle \bar{q}q \rangle \langle \bar{s}G s \rangle+6 \langle \bar{s}s \rangle \langle \bar{q}G q \rangle)}{6912 \pi ^4},\\
           \rho^{\langle\mathcal{O}_{12}\rangle}&=&\frac{s \left(1024 \pi ^4 \langle \bar{q}q \rangle^3 (\langle \bar{q}q \rangle+3 \langle \bar{s}s \rangle)\right)}{442368 \pi ^6}\nonumber\\
           &-&\frac{3s\langle GG \rangle (4 \langle \bar{q}q \rangle \langle \bar{q}G q \rangle+3 \langle \bar{q}q \rangle \langle \bar{s}G s \rangle+3 \langle \bar{s}s \rangle \langle \bar{q}G q \rangle)}{442368 \pi ^6},\\
              \rho^{\langle\mathcal{O}_{13}\rangle}&=&-\frac{m_s \langle GG \rangle^2 (6 \langle \bar{q}G q \rangle+\langle \bar{s}G s \rangle)}{452984832 \pi ^8}\nonumber\\
              &-&\frac{m_s  \langle \bar{q}G q \rangle(4 \langle \bar{q}q \rangle (9 \expval{\bar{q}G q}+\expval{\bar{s}G s})+3 \langle \bar{s}s \rangle \expval{\bar{q}G q})}{18432 \pi ^4}.
     \end{eqnarray}
   
   \subsubsection{Type-2 Current}
   \vspace{-1cm}
    \begin{eqnarray}
         \rho^{\text{pert}}&=&-\frac{s^7}{104044953600 \pi ^{10}},\\
          \rho^{\langle\mathcal{O}_3\rangle}&=&-\frac{m_s s^5 (7 \langle \bar{q}q \rangle+3 \langle \bar{s}s \rangle)}{206438400 \pi ^8},\\
          \rho^{\langle\mathcal{O}_4\rangle}&=&-\frac{\langle GG \rangle s^5}{1101004800 \pi ^{10}},\\
           \rho^{\langle\mathcal{O}_5\rangle}&=&\frac{m_s s^4 (18 \langle \bar{q}G q \rangle-5 \langle \bar{s}G s \rangle)}{70778880 \pi ^8},\\
            \rho^{\langle\mathcal{O}_6\rangle}&=&-\frac{\langle \bar{q}q \rangle s^4 (5 \langle \bar{q}q \rangle-3 \langle \bar{s}s \rangle)}{2211840 \pi ^6},\\
           \rho^{\langle\mathcal{O}_7\rangle}&=&-\frac{m_s \langle GG \rangle s^3 (5 \langle \bar{q}q \rangle+2 \langle \bar{s}s \rangle)}{5898240 \pi ^8},\\
           \rho^{\langle\mathcal{O}_8\rangle}&=&\frac{s^3 (16 \langle \bar{q}q \rangle \langle \bar{q}G q \rangle-5 \langle \bar{q}q \rangle \langle \bar{s}G s \rangle-5 \langle \bar{s}s \rangle \langle \bar{q}G q \rangle)}{737280 \pi ^6}-\frac{\langle GG \rangle^2 s^3}{4529848320 \pi ^{10}},\\
           \rho^{\langle\mathcal{O}_9\rangle}&=&\frac{m_s s^2 \left(3 \langle GG \rangle (4 \langle \bar{q}G q \rangle+\langle \bar{s}G s \rangle)-512 \pi ^4 \langle \bar{q}q \rangle^2 (8 \langle \bar{q}q \rangle+3 \langle \bar{s}s \rangle)\right)}{4718592 \pi ^8},\\
            \rho^{\langle\mathcal{O}_{10}\rangle}&=&\frac{s^2 \left(\langle \bar{q}q \rangle \langle GG \rangle (4 \langle \bar{s}s \rangle-3 \langle \bar{q}q \rangle)-9 \langle \bar{q}G q \rangle^2+6 \langle \bar{q}G q \rangle \langle \bar{s}G s \rangle\right)}{294912 \pi ^6},\\
            \rho^{\langle\mathcal{O}_{11}\rangle}&=&\frac{m_s \langle \bar{q}q \rangle s (27 \langle \bar{q}q \rangle \langle \bar{q}G q \rangle+2 \langle \bar{q}q \rangle \langle \bar{s}G s \rangle+6 \langle \bar{s}s \rangle \langle \bar{q}G q \rangle)}{6912 \pi ^4}\nonumber\\
            &-&\frac{m_s \langle GG \rangle^2 s (3 \langle \bar{q}q \rangle+\langle \bar{s}s \rangle)}{113246208 \pi ^8},\\
           \rho^{\langle\mathcal{O}_{12}\rangle}&=&\frac{ 3s \langle GG \rangle (4 \langle \bar{q}q \rangle \langle \bar{q}G q \rangle-3 \langle \bar{q}q \rangle \langle \bar{s}G s \rangle-3 \langle \bar{s}s \rangle \langle \bar{q}G q \rangle)}{442368 \pi ^6}\nonumber\\
           &-&\frac{s\left(1024 \pi ^4 \langle \bar{q}q \rangle^3 (\langle \bar{q}q \rangle-3 \langle \bar{s}s \rangle)\right)}{442368 \pi ^6},\\
              \rho^{\langle\mathcal{O}_{13}\rangle}&=&\frac{m_s \langle \bar{q}G q \rangle (-11 \expval{\bar{q}q} \langle \bar{q}G q \rangle+14 \expval{\bar{q}q} \langle \bar{s}G s \rangle+8 \langle \bar{s}s \rangle \langle \bar{q}G q \rangle)}{36864 \pi ^4}\nonumber\\
              &+&\frac{m_s \langle GG \rangle^2 (6 \langle \bar{q}G q \rangle+\langle \bar{s}G s \rangle)}{452984832 \pi ^8}.
     \end{eqnarray}
     
\section{The Spectral Densities of $p\bar\Sigma$ Hexaquark States}

    \subsection{$0^{-}$ $p\bar\Sigma$ Hexaquark States}
     \subsubsection{Type-1 Current}
     \vspace{-1cm}
     \begin{eqnarray}
         \rho^{\text{pert}}&=&\frac{s^7}{92484403200 \pi ^{10}},\\
          \rho^{\langle\mathcal{O}_3\rangle}&=&\frac{m_s \langle \bar{q}q \rangle s^5}{19660800 \pi ^8},\\
          \rho^{\langle\mathcal{O}_4\rangle}&=&\frac{\langle GG \rangle s^5}{943718400 \pi ^{10}},\\
           \rho^{\langle\mathcal{O}_5\rangle}&=&-\frac{m_s \langle \bar{q}G q \rangle s^4}{2949120 \pi ^8},\\
           \rho^{\langle\mathcal{O}_6\rangle}&=&\frac{\langle \bar{q}q \rangle s^4 (\langle \bar{s}s \rangle-2 \langle \bar{q}q \rangle)}{737280 \pi ^6},\\
           \rho^{\langle\mathcal{O}_7\rangle}&=&\frac{m_s \langle \bar{q}q \rangle \langle GG \rangle s^3}{1572864 \pi ^8},\\
           \rho^{\langle\mathcal{O}_8\rangle}&=&\frac{s^3 (4 \langle \bar{q}q \rangle \langle \bar{q}G q \rangle-\langle \bar{q}q \rangle \langle \bar{s}G s \rangle-\langle \bar{s}s \rangle \langle \bar{q}G q \rangle)}{147456 \pi ^6}+\frac{\langle GG \rangle^2 s^3}{3623878656 \pi ^{10}},\\
           \rho^{\langle\mathcal{O}_9\rangle}&=&\frac{m_s \langle \bar{q}q \rangle^2 s^2 (\langle \bar{s}s \rangle-\langle \bar{q}q \rangle)}{1536 \pi ^4}-\frac{m_s \langle GG \rangle \langle \bar{q}G q \rangle s^2}{589824 \pi ^8},\\
            \rho^{\langle\mathcal{O}_{10}\rangle}&=&\frac{\langle \bar{q}q \rangle \langle GG \rangle s^2 (\langle \bar{s}s \rangle-\langle \bar{q}q \rangle)}{73728 \pi ^6}+\frac{\langle \bar{q}G q \rangle s^2 (\langle \bar{s}G s \rangle-2 \langle \bar{q}G q \rangle)}{49152 \pi ^6},\\
            \rho^{\langle\mathcal{O}_{11}\rangle}&=&\frac{m_s \langle \bar{q}q \rangle s (26 \langle \bar{q}q \rangle \langle \bar{q}G q \rangle-9 \langle \bar{q}q \rangle \langle \bar{s}G s \rangle-17 \langle \bar{s}s \rangle \langle \bar{s}G s \rangle)}{9216 \pi ^4},\\
           \rho^{\langle\mathcal{O}_{12}\rangle}&=&\frac{\langle \bar{q}q \rangle^3 s (\langle \bar{q}q \rangle-2 \langle \bar{s}s \rangle)}{288 \pi ^2}+\frac{\langle GG \rangle s (2 \langle \bar{q}q \rangle \langle \bar{q}G q \rangle-\langle \bar{q}q \rangle \langle \bar{s}G s \rangle-\langle \bar{s}s \rangle \langle \bar{q}G q \rangle)}{49152 \pi ^6},\\
              \rho^{\langle\mathcal{O}_{13}\rangle}&=&\frac{m_s \langle \bar{q}G q \rangle (-25 \langle \bar{q}q \rangle \langle \bar{q}G q \rangle+17 \langle \bar{q}q \rangle \langle \bar{s}G s \rangle+8 \langle \bar{s}s \rangle \langle \bar{q}G q \rangle)}{18432 \pi ^4}.
     \end{eqnarray}

   \subsubsection{Type-2 Current}
   \vspace{-1cm}
    \begin{eqnarray}
         \rho^{\text{pert}}&=&\frac{s^7}{92484403200 \pi ^{10}},\\
          \rho^{\langle\mathcal{O}_3\rangle}&=&-\frac{m_s \langle \bar{q}q \rangle s^5}{58982400 \pi ^8},\\
          \rho^{\langle\mathcal{O}_4\rangle}&=&\frac{\langle GG \rangle s^5}{943718400 \pi ^{10}},\\
           \rho^{\langle\mathcal{O}_5\rangle}&=&\frac{m_s \langle \bar{q}G q \rangle s^4}{5898240 \pi ^8},\\
           \rho^{\langle\mathcal{O}_6\rangle}&=&\frac{\langle \bar{q}q \rangle s^4 (2 \langle \bar{q}q \rangle+\langle \bar{s}s \rangle)}{737280 \pi ^6},\\
           \rho^{\langle\mathcal{O}_7\rangle}&=&-\frac{m_s \langle \bar{q}q \rangle \langle GG \rangle s^3}{4718592 \pi ^8},\\
           \rho^{\langle\mathcal{O}_8\rangle}&=&\frac{\langle GG \rangle^2 s^3}{3623878656 \pi ^{10}}-\frac{s^3 (\langle \bar{q}q \rangle (4 \langle \bar{q}G q \rangle+\langle \bar{s}G s \rangle)+\langle \bar{s}s \rangle \langle \bar{q}G q \rangle)}{147456 \pi ^6},\\
           \rho^{\langle\mathcal{O}_9\rangle}&=&\frac{m_s \langle GG \rangle \langle \bar{q}G q \rangle s^2}{1179648 \pi ^8}-\frac{m_s \langle \bar{q}q \rangle^2 s^2 (\langle \bar{q}q \rangle+\langle \bar{s}s \rangle)}{4608 \pi ^4},\\
            \rho^{\langle\mathcal{O}_{10}\rangle}&=&\frac{\langle \bar{q}q \rangle \langle GG \rangle s^2 (\langle \bar{q}q \rangle+\langle \bar{s}s \rangle)}{73728 \pi ^6}+\frac{\langle \bar{q}G q \rangle s^2 (2 \langle \bar{q}G q \rangle+\langle \bar{s}G s \rangle)}{49152 \pi ^6},\\
            \rho^{\langle\mathcal{O}_{11}\rangle}&=&\frac{m_s \langle \bar{q}q \rangle s (10 \langle \bar{q}q \rangle \langle \bar{q}G q \rangle+3 \langle \bar{q}q \rangle \langle \bar{s}G s \rangle+7 \langle \bar{s}s \rangle \langle \bar{s}G s \rangle)}{9216 \pi ^4},\\
           \rho^{\langle\mathcal{O}_{12}\rangle}&=&\frac{\langle \bar{q}q \rangle^3 s (\langle \bar{q}q \rangle+2 \langle \bar{s}s \rangle)}{288 \pi ^2}-\frac{\langle GG \rangle s (\langle \bar{q}q \rangle (2 \langle \bar{q}G q \rangle+\langle \bar{s}G s \rangle)+\langle \bar{s}s \rangle \langle \bar{q}G q \rangle)}{49152 \pi ^6},\\
              \rho^{\langle\mathcal{O}_{13}\rangle}&=&-\frac{m_s \langle \bar{q}G q \rangle (11 \langle \bar{q}q \rangle \langle \bar{q}G q \rangle+7 \langle \bar{q}q \rangle \langle \bar{s}G s \rangle+4 \langle \bar{s}s \rangle \langle \bar{q}G q \rangle)}{18432 \pi ^4}.
     \end{eqnarray}

\subsection{$0^{+}$ $p\bar\Sigma$ Hexaquark States}
     \subsubsection{Type-1 Current}
     \vspace{-1cm}
     \begin{eqnarray}
         \rho^{\text{pert}}&=&\frac{s^7}{92484403200 \pi ^{10}},\\
          \rho^{\langle\mathcal{O}_3\rangle}&=&\frac{m_s \langle \bar{q}q \rangle s^5}{19660800 \pi ^8},\\
          \rho^{\langle\mathcal{O}_4\rangle}&=&\frac{\langle GG \rangle s^5}{943718400 \pi ^{10}},\\
           \rho^{\langle\mathcal{O}_5\rangle}&=&-\frac{m_s \langle \bar{q}G q \rangle s^4}{2949120 \pi ^8},\\
           \rho^{\langle\mathcal{O}_6\rangle}&=&-\frac{\langle \bar{q}q \rangle s^4 (2 \langle \bar{q}q \rangle+\langle \bar{s}s \rangle)}{737280 \pi ^6},\\
           \rho^{\langle\mathcal{O}_7\rangle}&=&\frac{m_s \langle \bar{q}q \rangle \langle GG \rangle s^3}{1572864 \pi ^8},\\
           \rho^{\langle\mathcal{O}_8\rangle}&=&\frac{s^3 (\langle \bar{q}q \rangle (4 \langle \bar{q}G q \rangle+\langle \bar{s}G s \rangle)+\langle \bar{s}s \rangle \langle \bar{q}G q \rangle)}{147456 \pi ^6}+\frac{\langle GG \rangle^2 s^3}{3623878656 \pi ^{10}},\\
           \rho^{\langle\mathcal{O}_9\rangle}&=&-\frac{m_s \langle \bar{q}q \rangle^2 s^2 (\langle \bar{q}q \rangle+\langle \bar{s}s \rangle)}{1536 \pi ^4}-\frac{m_s \langle GG \rangle \langle \bar{q}G q \rangle s^2}{589824 \pi ^8},\\
            \rho^{\langle\mathcal{O}_{10}\rangle}&=&-\frac{\langle \bar{q}q \rangle \langle GG \rangle s^2 (\langle \bar{q}q \rangle+\langle \bar{s}s \rangle)}{73728 \pi ^6}-\frac{\langle \bar{q}G q \rangle s^2 (2 \langle \bar{q}G q \rangle+\langle \bar{s}G s \rangle)}{49152 \pi ^6},\\
            \rho^{\langle\mathcal{O}_{11}\rangle}&=&\frac{m_s \langle \bar{q}q \rangle s (26 \langle \bar{q}q \rangle \langle \bar{q}G q \rangle+9 \langle \bar{q}q \rangle \langle \bar{s}G s \rangle+17 \langle \bar{s}s \rangle \langle \bar{s}G s \rangle)}{9216 \pi ^4},\\
           \rho^{\langle\mathcal{O}_{12}\rangle}&=&\frac{\langle \bar{q}q \rangle^3 s (\langle \bar{q}q \rangle+2 \langle \bar{s}s \rangle)}{288 \pi ^2}+\frac{\langle GG \rangle s (\langle \bar{q}q \rangle (2 \langle \bar{q}G q \rangle+\langle \bar{s}G s \rangle)+\langle \bar{s}s \rangle \langle \bar{q}G q \rangle)}{49152 \pi ^6},\\
              \rho^{\langle\mathcal{O}_{13}\rangle}&=&-\frac{m_s \langle \bar{q}G q \rangle (25 \langle \bar{q}q \rangle \langle \bar{q}G q \rangle+17 \langle \bar{q}q \rangle \langle \bar{s}G s \rangle+8 \langle \bar{s}s \rangle \langle \bar{q}G q \rangle)}{18432 \pi ^4}.
     \end{eqnarray}

   \subsubsection{Type-2 Current}
   \vspace{-1cm}
    \begin{eqnarray}
         \rho^{\text{pert}}&=&\frac{s^7}{92484403200 \pi ^{10}},\\
          \rho^{\langle\mathcal{O}_3\rangle}&=&-\frac{m_s \langle \bar{q}q \rangle s^5}{58982400 \pi ^8},\\
          \rho^{\langle\mathcal{O}_4\rangle}&=&\frac{\langle GG \rangle s^5}{943718400 \pi ^{10}},\\
           \rho^{\langle\mathcal{O}_5\rangle}&=&\frac{m_s \langle \bar{q}G q \rangle s^4}{5898240 \pi ^8},\\
           \rho^{\langle\mathcal{O}_6\rangle}&=&-\frac{\langle \bar{q}q \rangle s^4 (\langle \bar{s}s \rangle-2 \langle \bar{q}q \rangle)}{737280 \pi ^6},\\
           \rho^{\langle\mathcal{O}_7\rangle}&=&-\frac{m_s \langle \bar{q}q \rangle \langle GG \rangle s^3}{4718592 \pi ^8},\\
           \rho^{\langle\mathcal{O}_8\rangle}&=&-\frac{s^3 (4 \langle \bar{q}q \rangle \langle \bar{q}G q \rangle-\langle \bar{q}q \rangle \langle \bar{s}G s \rangle-\langle \bar{s}s \rangle \langle \bar{q}G q \rangle)}{147456 \pi ^6}+\frac{\langle GG \rangle^2 s^3}{3623878656 \pi ^{10}},\\
           \rho^{\langle\mathcal{O}_9\rangle}&=&-\frac{m_s \langle \bar{q}q \rangle^2 s^2 (\langle \bar{q}q \rangle-\langle \bar{s}s \rangle)}{4608 \pi ^4}+\frac{m_s \langle GG \rangle \langle \bar{q}G q \rangle s^2}{1179648 \pi ^8},\\
            \rho^{\langle\mathcal{O}_{10}\rangle}&=&-\frac{\langle \bar{q}q \rangle \langle GG \rangle s^2 (\langle \bar{s}s \rangle-\langle \bar{q}q \rangle)}{73728 \pi ^6}-\frac{\langle \bar{q}G q \rangle s^2 (\langle \bar{s}G s \rangle-2 \langle \bar{q}G q \rangle)}{49152 \pi ^6},\\
            \rho^{\langle\mathcal{O}_{11}\rangle}&=&\frac{m_s \langle \bar{q}q \rangle s (10 \langle \bar{q}q \rangle \langle \bar{q}G q \rangle-3 \langle \bar{q}q \rangle \langle \bar{s}G s \rangle-7 \langle \bar{s}s \rangle \langle \bar{s}G s \rangle)}{9216 \pi ^4},\\
           \rho^{\langle\mathcal{O}_{12}\rangle}&=&-\frac{\langle GG \rangle s (2 \langle \bar{q}q \rangle \langle \bar{q}G q \rangle-\langle \bar{q}q \rangle \langle \bar{s}G s \rangle-\langle \bar{s}s \rangle \langle \bar{q}G q \rangle)}{49152 \pi ^6}\\
           &+&\frac{\langle \bar{q}q \rangle^3 s (\langle \bar{q}q \rangle-2 \langle \bar{s}s \rangle)}{288 \pi ^2},\\
              \rho^{\langle\mathcal{O}_{13}\rangle}&=&-\frac{m_s \langle \bar{q}G q \rangle (11 \langle \bar{q}q \rangle \langle \bar{q}G q \rangle-7 \langle \bar{q}q \rangle \langle \bar{s}G s \rangle-4 \langle \bar{s}s \rangle \langle \bar{q}G q \rangle)}{18432 \pi ^4}.
     \end{eqnarray}

     \subsection{$1^{-}$ $p\bar\Sigma$ Hexaquark States}
     \subsubsection{Type-1 Current}
     \vspace{-1cm}
     \begin{eqnarray}
         \rho^{\text{pert}}&=&\frac{s^7}{104044953600 \pi ^{10}},\\
          \rho^{\langle\mathcal{O}_3\rangle}&=&\frac{m_s \langle \bar{q}q \rangle s^5}{22937600 \pi ^8},\\
          \rho^{\langle\mathcal{O}_4\rangle}&=&\frac{\langle GG \rangle s^5}{1101004800 \pi ^{10}},\\
           \rho^{\langle\mathcal{O}_5\rangle}&=&-\frac{m_s \langle \bar{q}G q \rangle s^4}{3538944 \pi ^8},\\
           \rho^{\langle\mathcal{O}_6\rangle}&=&-\frac{\langle \bar{q}q \rangle s^4 (5 \langle \bar{q}q \rangle-3 \langle \bar{s}s \rangle)}{2211840 \pi ^6},\\
           \rho^{\langle\mathcal{O}_7\rangle}&=&\frac{m_s \langle \bar{q}q \rangle \langle GG \rangle s^3}{1966080 \pi ^8},\\
           \rho^{\langle\mathcal{O}_8\rangle}&=&-\frac{s^3 (-16 \langle \bar{q}q \rangle \langle \bar{q}G q \rangle+5 \langle \bar{q}q \rangle \langle \bar{s}G s \rangle+5 \langle \bar{s}s \rangle \langle \bar{q}G q \rangle)}{737280 \pi ^6}+\frac{\langle GG \rangle^2 s^3}{4529848320 \pi ^{10}},\\
           \rho^{\langle\mathcal{O}_9\rangle}&=&-\frac{m_s \langle \bar{q}q \rangle^2 s^2 (3 \langle \bar{q}q \rangle-4 \langle \bar{s}s \rangle)}{6144 \pi ^4}-\frac{m_s \langle GG \rangle \langle \bar{q}G q \rangle s^2}{786432 \pi ^8},\\
            \rho^{\langle\mathcal{O}_{10}\rangle}&=&-\frac{\langle \bar{q}q \rangle \langle GG \rangle s^2 (3 \langle \bar{q}q \rangle-4 \langle \bar{s}s \rangle)}{294912 \pi ^6}-\frac{\langle \bar{q}G q \rangle s^2 (3 \langle \bar{q}G q \rangle-2 \langle \bar{s}G s \rangle)}{98304 \pi ^6},\\
            \rho^{\langle\mathcal{O}_{11}\rangle}&=&\frac{m_s \langle \bar{q}q \rangle s (52 \langle \bar{q}q \rangle \langle \bar{q}G q \rangle-27 \langle \bar{q}q \rangle \langle \bar{s}G s \rangle-51 \langle \bar{s}s \rangle \langle \bar{q}G q \rangle)}{27648 \pi ^4},\\
           \rho^{\langle\mathcal{O}_{12}\rangle}&=&-\frac{\langle GG \rangle s (-4 \langle \bar{q}q \rangle \langle \bar{q}G q \rangle+3 \langle \bar{q}q \rangle \langle \bar{s}G s \rangle+3 \langle \bar{s}s \rangle \langle \bar{q}G q \rangle)}{147456 \pi ^6}\\
           &+&\frac{\langle \bar{q}q \rangle^3 s (\langle \bar{q}q \rangle-3 \langle \bar{s}s \rangle)}{432 \pi ^2},\\
              \rho^{\langle\mathcal{O}_{13}\rangle}&=&\frac{m_s \langle \bar{q}G q \rangle (-25 \expval{\bar{q}q} \langle \bar{q}G q \rangle+34 \expval{\bar{q}q} \langle \bar{s}G s \rangle+16 \langle \bar{s}s \rangle \langle \bar{q}G q \rangle)}{36864 \pi ^4}
     \end{eqnarray}

   \subsubsection{Type-2 Current}
   \vspace{-1.5cm}
    \begin{eqnarray}
         \rho^{\text{pert}}&=&\frac{s^7}{104044953600 \pi ^{10}},\\
          \rho^{\langle\mathcal{O}_3\rangle}&=&-\frac{m_s \langle \bar{q}q \rangle s^5}{68812800 \pi ^8},\\
          \rho^{\langle\mathcal{O}_4\rangle}&=&\frac{\langle GG \rangle s^5}{1101004800 \pi ^{10}},\\
           \rho^{\langle\mathcal{O}_5\rangle}&=&\frac{m_s \langle \bar{q}G q \rangle s^4}{7077888 \pi ^8},\\
            \rho^{\langle\mathcal{O}_6\rangle}&=&\frac{\langle \bar{q}q \rangle s^4 (5 \langle \bar{q}q \rangle+3 \langle \bar{s}s \rangle)}{2211840 \pi ^6},\\
           \rho^{\langle\mathcal{O}_7\rangle}&=&-\frac{m_s \langle \bar{q}q \rangle \langle GG \rangle s^3}{5898240 \pi ^8},\\
           \rho^{\langle\mathcal{O}_8\rangle}&=&-\frac{s^3 (16 \langle \bar{q}q \rangle \langle \bar{q}G q \rangle+5 \langle \bar{q}q \rangle \langle \bar{s}G s \rangle+5 \langle \bar{s}s \rangle \langle \bar{q}G q \rangle)}{737280 \pi ^6}+\frac{\langle GG \rangle^2 s^3}{4529848320 \pi ^{10}},\\
           \rho^{\langle\mathcal{O}_9\rangle}&=&-\frac{m_s \langle \bar{q}q \rangle^2 s^2 (3 \langle \bar{q}q \rangle+4 \langle \bar{s}s \rangle)}{18432 \pi ^4}+\frac{m_s \langle GG \rangle \langle \bar{q}G q \rangle s^2}{1572864 \pi ^8},\\
            \rho^{\langle\mathcal{O}_{10}\rangle}&=&\frac{\langle \bar{q}q \rangle \langle GG \rangle s^2 (3 \langle \bar{q}q \rangle+4 \langle \bar{s}s \rangle)}{294912 \pi ^6}+\frac{\langle \bar{q}G q \rangle s^2 (3 \langle \bar{q}G q \rangle+2 \langle \bar{s}G s \rangle)}{98304 \pi ^6},\\
            \rho^{\langle\mathcal{O}_{11}\rangle}&=&\frac{m_s \langle \bar{q}q \rangle s (20 \langle \bar{q}q \rangle \langle \bar{q}G q \rangle+9 \langle \bar{q}q \rangle \langle \bar{s}G s \rangle+21 \langle \bar{s}s \rangle \langle \bar{q}G q \rangle)}{27648 \pi ^4},\\
           \rho^{\langle\mathcal{O}_{12}\rangle}&=&-\frac{\langle GG \rangle s (4 \langle \bar{q}q \rangle \langle \bar{q}G q \rangle+3 \langle \bar{q}q \rangle \langle \bar{s}G s \rangle+3 \langle \bar{s}s \rangle \langle \bar{q}G q \rangle)}{147456 \pi ^6}\\
           &+&\frac{\langle \bar{q}q \rangle^3 s (\langle \bar{q}q \rangle+3 \langle \bar{s}s \rangle)}{432 \pi ^2},\\
              \rho^{\langle\mathcal{O}_{13}\rangle}&=&-\frac{m_s \langle \bar{q}G q \rangle (11 \expval{\bar{q}q} \langle \bar{q}G q \rangle+14 \expval{\bar{q}q} \langle \bar{s}G s \rangle+8 \langle \bar{s}s \rangle \langle \bar{q}G q \rangle)}{36864 \pi ^4}.
     \end{eqnarray}

\subsection{$1^{+}$ $p\bar\Sigma$ Hexaquark States}
     \subsubsection{Type-1 Current}
     \vspace{-1.5cm}
     \begin{eqnarray}
         \rho^{\text{pert}}&=&\frac{s^7}{104044953600 \pi ^{10}},\\
          \rho^{\langle\mathcal{O}_3\rangle}&=&\frac{m_s s^5 (16 \langle \bar{q}q \rangle+3 \langle \bar{s}s \rangle)}{206438400 \pi ^8},\\
          \rho^{\langle\mathcal{O}_4\rangle}&=&\frac{\langle GG \rangle s^5}{1101004800 \pi ^{10}},\\
           \rho^{\langle\mathcal{O}_5\rangle}&=&-\frac{m_s s^4 (38 \langle \bar{q}G q \rangle+5 \langle \bar{s}G s \rangle)}{70778880 \pi ^8},\\
           \rho^{\langle\mathcal{O}_6\rangle}&=&-\frac{\langle \bar{q}q \rangle s^3 \left(s (5 \langle \bar{q}q \rangle+3 \langle \bar{s}s \rangle)-18 m_s^2 (5 \langle \bar{q}q \rangle+2 \langle \bar{s}s \rangle)\right)}{2211840 \pi ^6},\\
           \rho^{\langle\mathcal{O}_7\rangle}&=&\frac{m_s \langle GG \rangle s^3 (4 \langle \bar{q}q \rangle+\langle \bar{s}s \rangle)}{2949120 \pi ^8},\\
           \rho^{\langle\mathcal{O}_8\rangle}&=&-\frac{17 m_s^2 \langle \bar{q}q \rangle \langle \bar{q}Gq \rangle s^2}{73728 \pi ^6}-\frac{m_s^2 \langle \bar{q}q \rangle \langle \bar{s}Gs \rangle s^2}{32768 \pi ^6}-\frac{m_s^2 \langle \bar{s}s \rangle \langle \bar{q}Gq \rangle s^2}{24576 \pi ^6}\nonumber \\
           &\phantom{=}&+\frac{\langle \bar{q}q \rangle \langle \bar{q}Gq \rangle s^3}{46080 \pi ^6}+\frac{\langle \bar{q}q \rangle \langle \bar{s}Gs \rangle s^3}{147456 \pi ^6}+\frac{\langle \bar{s}s \rangle \langle \bar{q}Gq \rangle s^3}{147456 \pi ^6}+\frac{\langle GG \rangle^2 s^3}{4529848320 \pi ^{10}},\\
           \rho^{\langle\mathcal{O}_9\rangle}&=&-\frac{m_s s^2 \left(256 \pi ^4 \langle \bar{q}q \rangle^2 (25 \langle \bar{q}q \rangle+18 \langle \bar{s}s \rangle)+3 \langle GG \rangle (6 \langle \bar{q}G q \rangle+\langle \bar{s}G s \rangle)\right)}{4718592 \pi ^8},\\
            \rho^{\langle\mathcal{O}_{10}\rangle}&=&-\frac{\langle \bar{q}q \rangle \langle GG \rangle s \left(s (3 \langle \bar{q}q \rangle+4 \langle \bar{s}s \rangle)-6 m_s^2 (3 \langle \bar{q}q \rangle+\langle \bar{s}s \rangle)\right)}{294912 \pi ^6}\nonumber\\
            &\phantom{=}&-\frac{\langle \bar{q}Gq \rangle s \left(9 s (3 \langle \bar{q}Gq \rangle+2 \langle \bar{s}Gs \rangle)-16 m_s^2 (9 \langle \bar{q}Gq \rangle+2 \langle \bar{s}Gs \rangle)\right)}{884736 \pi ^6},\\
            \rho^{\langle\mathcal{O}_{11}\rangle}&=&\frac{m_s \langle GG \rangle^2 s (3 \langle \bar{q}q \rangle+\langle \bar{s}s \rangle)}{113246208 \pi ^8}\nonumber\\
            &\phantom{=}&+\frac{5 m_s \langle \bar{q}q \rangle s (32 \langle \bar{q}q \rangle \langle \bar{q}G q \rangle+7 \langle \bar{q}q \rangle \langle \bar{s}G s \rangle+15 \langle \bar{s}s \rangle \langle \bar{q}G q \rangle)}{27648 \pi ^4},\\
           \rho^{\langle\mathcal{O}_{12}\rangle}&=&-\frac{\langle \bar{q}q \rangle^3 \left(9 m_s^2 (4 \langle \bar{q}q \rangle+\langle \bar{s}s \rangle)-8 s (\langle \bar{q}q \rangle+3 \langle \bar{s}s \rangle)\right)}{3456 \pi ^2}\nonumber\\
           &\phantom{=}&-\frac{\langle GG \rangle m_s^2 (34 \langle \bar{q}q \rangle \langle \bar{q}Gq \rangle+3 \langle \bar{q}q \rangle \langle \bar{s}Gs \rangle+4 \langle \bar{s}s \rangle \langle \bar{q}Gq \rangle)}{589824 \pi ^6}\nonumber\\
           &\phantom{=}&+\frac{4 s\langle GG \rangle (4 \langle \bar{q}q \rangle \langle \bar{q}Gq \rangle+3 \langle \bar{q}q \rangle \langle \bar{s}Gs \rangle+3 \langle \bar{s}s \rangle \langle \bar{q}Gq \rangle)}{589824 \pi ^6},\\
              \rho^{\langle\mathcal{O}_{13}\rangle}&=&-\frac{m_s \langle GG \rangle^2 (6 \langle \bar{q}G q \rangle+\langle \bar{s}G s \rangle)}{452984832 \pi ^8}\nonumber\\
              &\phantom{=}&-\frac{m_s \langle \bar{q}G q \rangle (97 \expval{\bar{q}q} \langle \bar{q}G q \rangle+42 \expval{\bar{q}q} \langle \bar{s}G s \rangle+22 \langle \bar{s}s \rangle \langle \bar{q}G q \rangle)}{36864 \pi ^4}
     \end{eqnarray}

   \subsubsection{Type-2 Current}
   \vspace{-1.5cm}
    \begin{eqnarray}
         \rho^{\text{pert}}&=&-\frac{s^7}{104044953600 \pi ^{10}},\\
          \rho^{\langle\mathcal{O}_3\rangle}&=&\frac{m_s \langle \bar{q}q \rangle s^5}{68812800 \pi ^8},\\
          \rho^{\langle\mathcal{O}_4\rangle}&=&-\frac{\langle GG \rangle s^5}{1101004800 \pi ^{10}},\\
           \rho^{\langle\mathcal{O}_5\rangle}&=&-\frac{m_s \langle \bar{q}G q \rangle s^4}{7077888 \pi ^8},\\
            \rho^{\langle\mathcal{O}_6\rangle}&=&-\frac{\langle \bar{q}q \rangle s^4 (5 \langle \bar{q}q \rangle-3 \langle \bar{s}s \rangle)}{2211840 \pi ^6},\\
           \rho^{\langle\mathcal{O}_7\rangle}&=&\frac{m_s \langle \bar{q}q \rangle \langle GG \rangle s^3}{5898240 \pi ^8},\\
           \rho^{\langle\mathcal{O}_8\rangle}&=&\frac{s^3 (16 \langle \bar{q}q \rangle \langle \bar{q}G q \rangle-5 \langle \bar{q}q \rangle \langle \bar{s}G s \rangle-5 \langle \bar{s}s \rangle \langle \bar{q}G q \rangle)}{737280 \pi ^6}-\frac{\langle GG \rangle^2 s^3}{4529848320 \pi ^{10}},\\
           \rho^{\langle\mathcal{O}_9\rangle}&=&\frac{m_s s^2 \left(256 \pi ^4 \langle \bar{q}q \rangle^2 (3 \langle \bar{q}q \rangle-4 \langle \bar{s}s \rangle)-3 \langle GG \rangle \langle \bar{q}G q \rangle\right)}{4718592 \pi ^8},\\
            \rho^{\langle\mathcal{O}_{10}\rangle}&=&\frac{s^2 \left(\langle \bar{q}q \rangle \langle GG \rangle (4 \langle \bar{s}s \rangle-3 \langle \bar{q}q \rangle)-9 \langle \bar{q}G q \rangle^2+6 \langle \bar{q}G q \rangle \langle \bar{s}G s \rangle\right)}{294912 \pi ^6},\\
            \rho^{\langle\mathcal{O}_{11}\rangle}&=&-\frac{m_s \langle \bar{q}q \rangle s (20 \langle \bar{q}q \rangle \langle \bar{q}G q \rangle-9 \langle \bar{q}q \rangle \langle \bar{s}G s \rangle-21 \langle \bar{s}s \rangle \langle \bar{q}G q \rangle)}{27648 \pi ^4},\\
           \rho^{\langle\mathcal{O}_{12}\rangle}&=&\frac{3s\langle GG \rangle (4 \langle \bar{q}q \rangle \langle \bar{q}G q \rangle-3 \langle \bar{q}q \rangle \langle \bar{s}G s \rangle-3 \langle \bar{s}s \rangle \langle \bar{q}G q \rangle)}{442368 \pi ^6}\nonumber\\
           &\phantom{=}&-\frac{1024 \pi ^4 \langle \bar{q}q \rangle^3 (\langle \bar{q}q \rangle-3 \langle \bar{s}s \rangle)}{432 \pi ^6},\\
              \rho^{\langle\mathcal{O}_{13}\rangle}&=&\frac{m_s \langle \bar{q}G q \rangle (-11 \expval{\bar{q}q} \langle \bar{q}G q \rangle+14 \expval{\bar{q}q} \langle \bar{s}G s \rangle+8 \langle \bar{s}s \rangle \langle \bar{q}G q \rangle)}{36864 \pi ^4}.
     \end{eqnarray}

\end{appendix}

\end{document}